\documentclass[iop,apjl,numberedappendix]{emulateapj}
\usepackage{natbib}
\usepackage{graphicx}
\usepackage{amsmath,amsfonts}
\usepackage{booktabs}
\usepackage{hyperref}
\usepackage{afterpage} 
\usepackage[normalem]{ulem}

\hypersetup{colorlinks=false,pdfborder={0 0 0}}

\slugcomment{Accepted for publication in ApJ, October 16, 2016}

\setcitestyle{notesep={; }}

\def\sgra{Sgr~A$^{\ast}$}

\def\lsim{\mathrel{\raise.3ex\hbox{$<$\kern-.75em\lower1ex\hbox{$\sim$}}}}
\def\gsim{\mathrel{\raise.3ex\hbox{$>$\kern-.75em\lower1ex\hbox{$\sim$}}}}
\def\gtwid{\mathrel{\raise.3ex\hbox{$>$\kern-.75em\lower1ex\hbox{$\sim$}}}}
\def\proptwid{\mathrel{\raise.3ex\hbox{$\propto$\kern-.75em\lower1ex\hbox{$\sim$}}}}

\begin{document}

\title{ Stochastic Optics:\\\vspace{0.5ex}A Scattering Mitigation Framework for Radio Interferometric Imaging }
\shorttitle{Stochastic Optics}

\author{Michael D.~Johnson}
\shortauthors{Michael D.~Johnson}
\affil{Harvard-Smithsonian Center for Astrophysics, 60 Garden Street, Cambridge, MA 02138, USA}
\email{mjohnson@cfa.harvard.edu} 


\begin{abstract}
Just as turbulence in the Earth's atmosphere can severely limit the angular resolution of optical telescopes, turbulence in the ionized interstellar medium fundamentally limits the resolution of radio telescopes. We present a scattering mitigation framework for radio imaging with very long baseline interferometry (VLBI) that partially overcomes this limitation. Our framework, ``stochastic optics,''  derives from a simplification of strong interstellar scattering to separate small-scale (``diffractive'') effects from large-scale (``refractive'') effects, thereby separating deterministic and random contributions to the scattering. Stochastic optics extends traditional synthesis imaging by simultaneously reconstructing an unscattered image and its refractive perturbations. Its advantages over direct imaging come from utilizing the many deterministic properties of the scattering -- such as the time-averaged ``blurring,'' polarization independence, and the deterministic evolution in frequency and time -- while still accounting for the stochastic image distortions on large scales. These distortions are identified in the image reconstructions through regularization by their time-averaged power spectrum. Using synthetic data, we show that this framework effectively removes the blurring from diffractive scattering while reducing the spurious image features from refractive scattering. Stochastic optics can provide significant improvements over existing scattering mitigation strategies and is especially promising for imaging the Galactic Center supermassive black hole, Sagittarius~A$^\ast$, with the Global mm-VLBI Array and with the Event Horizon Telescope. 
\end{abstract}

\section{Introduction}

For its entire history, optical astronomy has had to contend with the deleterious effects of the Earth's atmosphere on images. These effects are apparent even with the naked eye through the twinkling of stars. A major technological advance, first proposed by \citet{Babcock_1953}, was the implementation of adaptive optics to dynamically correct for the rapid wavefront variations caused by the atmosphere, allowing optical and infrared telescopes to approach their diffraction-limited resolutions \citep[reviewed by][]{Davies_Kasper_2012}. The effects of the Earth's atmosphere on radio observations with a narrow field of view are simpler, introducing a single time- and frequency-dependent phase for a telescope, which can be removed as part of standard calibration \citep{Pearson_Readhead_1984,TMS}. But radio observations are subject to another source of random, time-variable refraction -- the ionized interstellar medium (ISM). 

Both atmospheric and interstellar scattering can be described in the framework of wave propagation through an irregular refracting medium. And both are typically analyzed in the simplified framework of a thin scattering screen that only modifies the phase of the incident radiation \citep[see, e.g.,][]{Narayan_1992}. A major difference, however, is that optical scattering in the atmosphere corresponds to a screen whose phase varies by much less than 1 radian across the scattered image, while radio-wave scattering in the ISM often has phase variations of many radians across the scattered image. These two cases describe the weak and strong scattering regimes, respectively. 

In the weak-scattering regime, the dominant effects of scattering are from large-scale phase gradients, which primarily cause the image to become speckled and its centroid to shift randomly with time. Indeed, many mitigation strategies for optical scattering entirely focus on removing the image wander (in adaptive optics, a so-called ``tip-tilt'' correction). In the strong-scattering regime, the scattering effects bifurcate into two classes: ``diffractive'' and ``refractive''. Diffractive effects arise from small-scale phase gradients; their most familiar effect on radio images is ``blurring'' with a kernel that grows approximately with the squared observing wavelength. Refractive effects arise from large-scale phase gradients; they produce image distortions and introduce substructure into the scattered image \citep{NarayanGoodman89,GoodmanNarayan89,Johnson_Gwinn_2015}. With the advent of microarcsecond imaging with very long baseline interferometry (VLBI), signatures of refractive substructure are now apparent in observations and can significantly affect VLBI imaging \citep[e.g.,][]{Gwinn_2014,Johnson_2016,Ortiz_2016}. As discussed in \citet{Johnson_Gwinn_2015}, refractive effects are especially important for imaging the Galactic Center supermassive black hole, Sagittarius~A$^\ast$ (\sgra), with the Event Horizon Telescope \citep[EHT;][]{Doeleman_2009} and for imaging active galactic nuclei (AGN) using Earth-space VLBI with {\it RadioAstron} \citep{Kardashev_2013}. 

In this paper, we develop a new scattering mitigation strategy for VLBI imaging, ``stochastic optics''. This strategy derives from the separation of diffractive and refractive scattering effects, approximating the former by their time-averaged image blurring while retaining the latter through a stochastic, large-scale scattering screen \citep{Blandford_Narayan_1985,Johnson_Narayan_2016}. Stochastic optics extends traditional VLBI synthesis imaging by simultaneously reconstructing the unscattered image and the refractive scattering screen. It is analogous to adaptive optics but does not require a natural or artificial guide star (i.e., a bright point source), instead utilizing known statistical properties of the scattering medium to regularize the image reconstruction. We describe the theoretical motivation for our approach in \S\ref{sec::Background}. We then outline how to implement stochastic optics in VLBI imaging and show example image reconstructions with synthetic data in \S\ref{sec::Mitigation_Framework}. We discuss the relationship between stochastic optics and existing mitigation strategies in \S\ref{sec::Discussion}, and we summarize our method and conclusions in \S\ref{sec::Summary}.

\section{Background}
\label{sec::Background}

The motivation and framework of thin-screen scattering have been extensively developed and are summarized in several excellent reviews \citep[e.g.,][]{Rickett_1990,Narayan_1992,TMS}. In this section, we only provide a brief overview of scattering theory that is focused on the aspects that are relevant to our mitigation strategy. 

\subsection{Thin-Screen Scattering}
\label{sec::Thin-Screen}

When an image is viewed through a medium with spatial variations in its refractive index, the image is distorted. Refractive inhomogeneities steer and focus different regions of an image while preserving surface brightness \citep{Born_Wolf}. For optical images, familiar examples of such distortions arise from peering through a medium with steep temperature gradients -- for instance, an image viewed above a flame. In contrast, radio-wave scattering in the ionized ISM arises from density inhomogeneities because the refractivity in a plasma is approximately proportional to the local electron density \citep{Jackson_1999}. 

For both optical and radio observations, the scattering is often well-described as arising from a turbulent cascade that is localized to a single thin screen between the source and observer. The screen adds a stochastic, position-dependent phase $\phi(\mathbf{r})$ to the incident radiation but does not alter the amplitude of incident waves. Here, and throughout this paper, $\mathbf{r}$ denotes a two-dimensional transverse coordinate on the screen. The screen is assumed to be statistically homogeneous, and the scattering can then be quantified in two complementary ways: by the structure function of the phase fluctuations $D_{\phi}(\mathbf{r}) \equiv \left \langle \left[ \phi\left( \mathbf{r}' + \mathbf{r} \right) - \phi\left( \mathbf{r'} \right) \right]^2 \right \rangle \propto \left| \mathbf{r} \right|^\alpha$ or by the power spectrum of the phase fluctuations $Q(\mathbf{q}) \propto \left| \mathbf{q} \right|^{-(\alpha+2)}$. These two approaches are related by a Fourier transform, $Q(\mathbf{q}) = -\frac{1}{2\lambdabar^2} \tilde{D}_{\phi}(\mathbf{q})$, where the prefactor renders $Q(\mathbf{q})$ dimensionless and independent of the observing wavelength ($\lambdabar \equiv \frac{\lambda}{2\pi}$). We will adopt a Kolmogorov index ($\alpha=5/3$) for examples, as is observationally motivated \citep{Armstrong_1995}, although our mitigation strategy is suitable for any ``shallow'' spectrum ($0 < \alpha < 2$). 

In addition to $\alpha$, a small number of parameters characterize the scattering. The first is the \emph{phase coherence length} $r_0$, defined by $D_{\phi}\left(r_0\right) = 1$ \citep[$r_0$ is closely related to the {\it Fried parameter} in the optical literature; see][]{Narayan_1992}. The second is the \emph{Fresnel scale} $r_{\rm F} \equiv \sqrt{ \frac{ D R }{D + R} \lambdabar }$, where $D$ is the Earth-scattering distance, and $R$ is the source-scattering distance. These parameters also determine the scattering regime: when $r_0 < r_{\rm F}$ the scattering is ``strong,'' and when $r_0 > r_{\rm F}$ the scattering is ``weak.'' Our scattering mitigation strategy will apply to observations in the strong scattering regime.  The evolution of the scattering in time is most commonly estimated using the frozen-screen approximation \citep{Taylor_1938}, which depends on a characteristic transverse velocity vector $\mathbf{V}_\perp$. The most common extensions to this basic scattering model are to include inner and outer scales of the turbulence and anisotropic scattering. 

Other useful quantities can then be written in terms of these parameters. For instance, the \emph{refractive scale}, $r_{\rm R} = r_{\rm F}^2/r_0$ determines the transverse size of the ensemble-average scattered image of a point source. The full width at half maximum (FWHM) angular size of this image is $\theta_{\rm scatt} \approx 0.37 (1+M)^{-1}\lambda/r_0 = 0.37 (2\pi r_{\rm R}/D)$, where $M \equiv D/R$ is the effective magnification of the scattering screen when viewed as a lens \citep[see][]{Gwinn_1998,Johnson_Gwinn_2015}. In the optical literature, this image is called the \emph{seeing disk}, and $\theta_{\rm scatt}$ is called the \emph{seeing}. The \emph{isoplanatic angle} $\theta_{\rm iso}$ describes the angular displacement over which a point source will have a similar instantaneous scattered image. For the optical case of weak scintillation, $\theta_{\rm iso} \sim r_0/D \gg \theta_{\rm scatt}$, whereas for the radio case of refractive scattering, $\theta_{\rm iso} \sim r_{\rm R}/D = \theta_{\rm scatt}$.  

Despite their similar scattering foundations, optical scattering in the atmosphere and radio scattering in the ionized ISM have key differences. For example, the atmospheric refractivity for wavelengths from optical through infrared is only weakly dependent on wavelength \citep{Edlen_1966}, so optical scattering is nearly achromatic. Consequently, $\phi \propto \lambda^{-1}$, $r_0 \propto \lambda^{6/5}$, and $r_{\rm R} \propto \lambda^{-1/5}$. In contrast, the refractivity for radio-wave propagation in the ionized ISM (a cold plasma) is proportional to wavelength \citep{Jackson_1999}, so $\phi \propto \lambda$, $r_0 \propto \lambda^{-6/5}$, and $r_{\rm R} \propto \lambda^{11/5}$. Another difference is that optical scattering in the atmosphere is typically weak \citep[see][]{Narayan_1992}, while radio scattering is typically strong below ${\sim}5\,{\rm GHz}$; for heavily scattered lines of sight, the strong-weak transition frequency is higher (e.g., it is ${\sim}2\,{\rm THz}$ for \sgra).  A third important difference is that optical and radio scattering have vastly different coherence timescales: milliseconds for optical observations and days to months for radio observations. These different timescales result in different challenges and strategies for mitigation, as we will discuss later in \S\ref{sec::AO_Comparison}.

\subsection{Assessing Scattering Degradation: The Strehl Ratio}

\begin{figure}[t]
\centering
\includegraphics[width=0.48\textwidth]{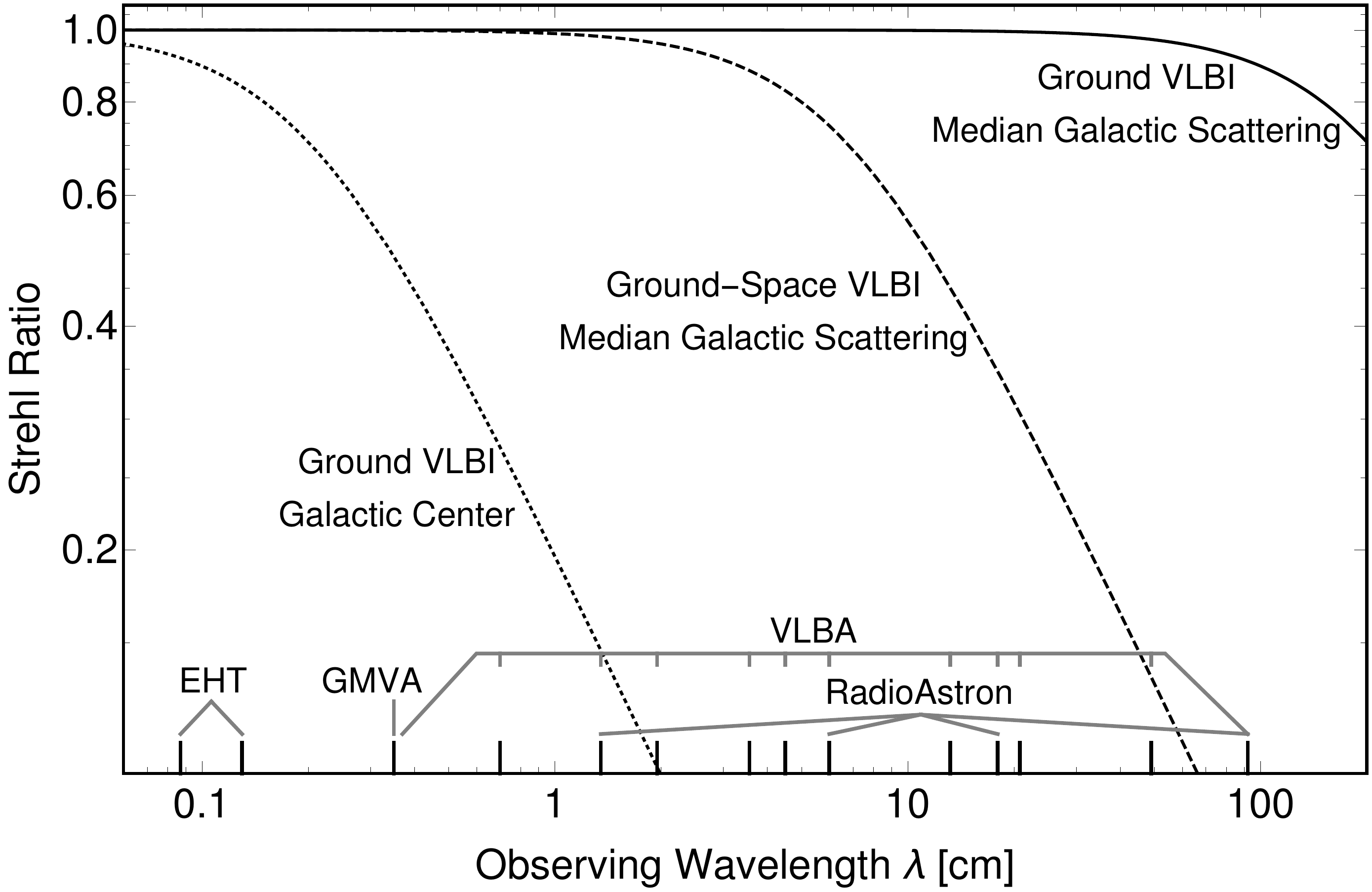}
\caption
{ 
The Strehl ratio (Eq.~\ref{eq::Radio_Strehl}) as a function of observing wavelength for three scenarios: (solid) ground-based VLBI with median Galactic scattering, (dashed) ground-space VLBI with median Galactic scattering, and (dotted) ground-based VLBI with scattering corresponding to the Galactic center. For ground-space VLBI, we assume baselines extending to 30 Earth diameters, similar to the range of {\it RadioAstron}. Heavy ticks along the wavelength axis indicate operating wavelengths of several current arrays. Mitigation strategies will be important for observations at 6- and 18-cm with {\it RadioAstron} and for ground-based observations of \sgra\ at wavelengths even as short as 1\,mm.
}
\label{fig::Strehl_Ratio}
\end{figure}

The \emph{Strehl ratio}, $0 < S \leq 1$, is commonly used in the optical literature to characterize image degradation from scattering. $S$ is defined as the peak intensity of the measured point spread function (with scattering) divided by the peak intensity of the ideal, diffraction-limited instrumental response (without scattering). 

We can extend the Strehl ratio to radio interferometry by relating the peak brightness of an unscattered image restored with the nominal VLBI beam (FWHM $\theta_{\rm uv}$; hereafter the ``CLEAN beam'') to the peak brightness of the ensemble-average scattered image restored with the same beam (approximate FWHM $\sqrt{\theta_{\rm uv}^2 + \theta_{\rm scatt}^2}$). For Gaussian images, the peak brightness is proportional to the inverse FWHM, so we obtain
\begin{align}
\label{eq::Radio_Strehl}
S &\approx \frac{\theta_{\rm uv}}{\sqrt{\theta_{\rm uv}^2 + \theta_{\rm scatt}^2}}.
\end{align}
Thus, the Strehl ratio gives the angular resolution that a VLBI array achieves in the presence of scattering as a fraction of its nominal diffraction limit. 

In the optical community, the Strehl ratio is used to quantify the significance of scattering and also to assess the performance of scattering mitigation techniques. For the radio case, only the former application is relevant because we will see that existing scattering mitigation techniques tend to introduce spurious compact features in reconstructed images (corresponding to $S>1$). 

We will now estimate the Strehl ratio for a few cases of interest. 
A VLBI array spanning the Earth has maximal baseline lengths of ${\sim}10^4\,{\rm km}$, so $\theta_{\rm uv} \approx \lambda_{\rm cm} \times (0.2\ {\rm mas})$, where $\lambda_{\rm cm}$ is the observing wavelength in cm.\footnote{Note that $\theta_{\rm uv}$ in this case corresponds to the nominal resolution of a global VLBI array; some imaging techniques regularly achieve a factor of ${\approx}2$ improvement \citep[e.g.,][]{Cornwell_Evans_1985,Chael_2016}, resulting in a lower Strehl ratio.} For typical lines of sight away from the Galactic plane, $\theta_{\rm scatt} \sim \lambda_{\rm cm}^2 \times (1\, \mu{\rm as})$ \citep{NE2001,Johnson_Gwinn_2015}. These characteristic parameters give a Strehl ratio near unity for wavelengths shorter than a decimeter, and $S \sim 200/\lambda_{\rm cm}$ for wavelengths longer than a few decimeters. However, for Earth-space VLBI with {\it RadioAstron}, the Strehl ratio is near unity only for wavelengths shorter than ${\sim}5\,{\rm cm}$, falling as $S \sim 7/\lambda_{\rm cm}$ for longer wavelengths. For heavily scattered lines of sight, such as those near the Galactic plane, $S$ falls below unity at much shorter wavelengths. For instance,  the line of sight to the Galactic center has $\theta_{\rm scatt} \sim \lambda_{\rm cm}^2 \times (1\, {\rm mas})$ \citep[see][]{Bower_2006}, giving $S \approx 0.2/\lambda_{\rm cm}$ for Earth-based VLBI at wavelengths longer than a few millimeters. Figure~\ref{fig::Strehl_Ratio} shows the Strehl ratio as a function of observing wavelength for these cases.

\begin{figure*}[t]
\centering
\includegraphics[width=0.705\textwidth]{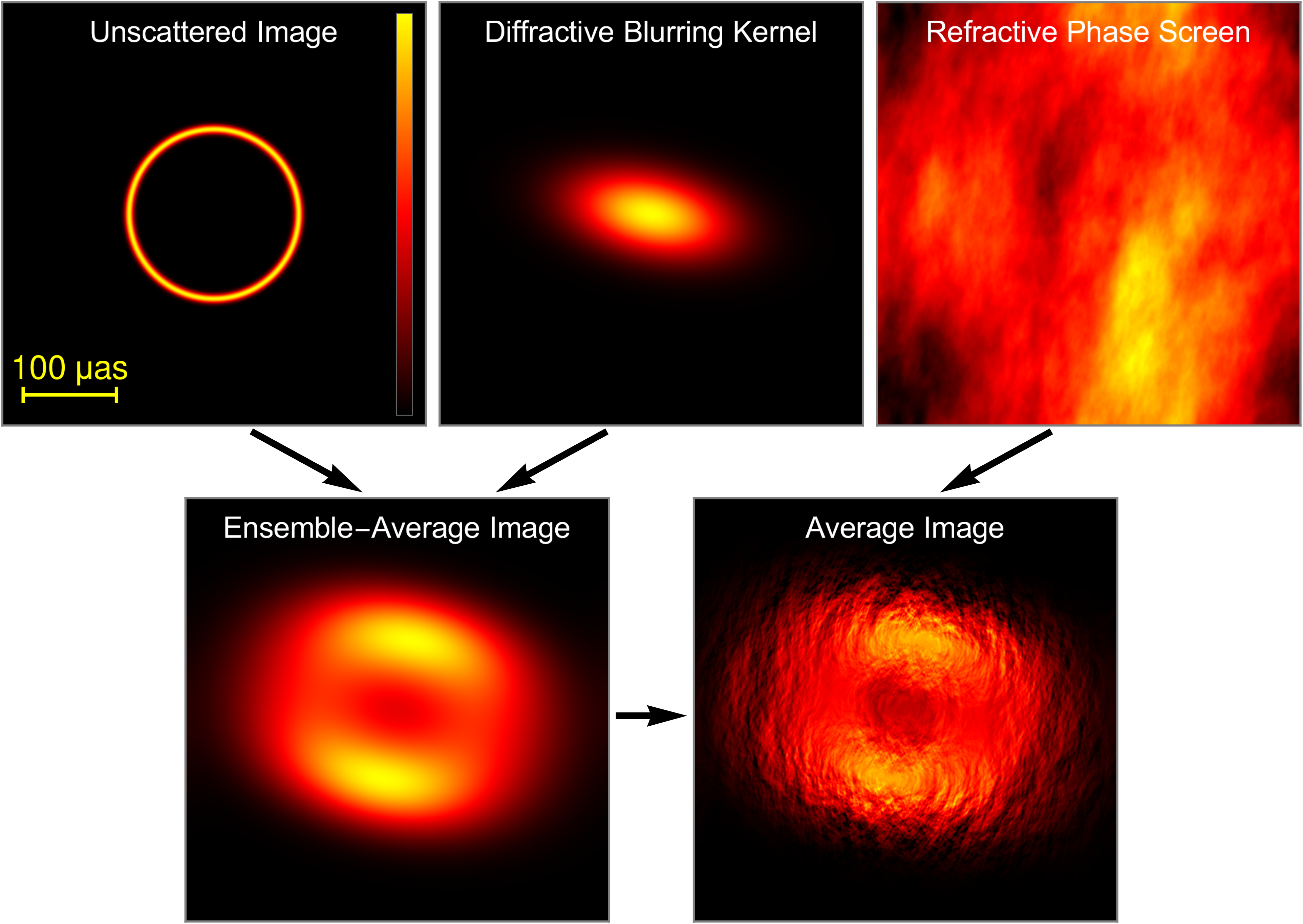}
\caption
{ 
Relationships between different scattering and imaging regimes in the geometric optics reduction of \citet{Blandford_Narayan_1985}. For simplicity, the unscattered model is a thin ring of radius $100\,\mu{\rm as}$ that has a $5\,\mu{\rm as}$ Gaussian taper in the radial direction (this ring is not associated with the black hole shadow of \sgra); the scattering parameters correspond to those of \sgra\ at 3.5\,mm (see \S\ref{sec::Example_Reconstructions}). As discussed in \S\ref{sec::DiffractiveRefractive}, this reduction decouples scattering into diffractive and refractive contributions, which give rise to deterministic and stochastic image distortions, respectively. The ensemble-average image is the convolution of the unscattered image with the diffractive kernel. The average image is a distorted version of the ensemble-average image after steering and focusing of rays by gradients of the refractive phase screen (Eq.~\ref{eq::Geometric_Optics}). The color scale in each frame is linear and each is scaled independently; the total range of screen phase in this example spans $1.4 \times 10^4~{\rm radians}$.   Stochastic optics uses observations of the average image along with the known kernel $\tilde{G}(\mathbf{b})$ and (time-averaged) power spectrum $Q(\mathbf{q})$ of the scattering to simultaneously reconstruct the unscattered image and the refractive phase screen. 
}
\label{fig::Imaging_Overview}
\end{figure*}

\subsection{Decoupling Diffractive and Refractive Scattering}
\label{sec::DiffractiveRefractive}

In weak scattering, the incident wavefront is only mildly perturbed, and the dominant scattering effects arise from modes in the phase screen with $|\mathbf{q}| \sim r_{\rm F}^{-1}$. In strong scattering, the dominant effects arise from fluctuations on two widely separated scales $r_0$ and $r_{\rm R}$, giving rise to two branches of scintillation, diffractive and refractive \citep{RCB_1984}. \citet{NarayanGoodman89} and \citet{GoodmanNarayan89} showed that there are then three applicable averaging regimes for images. The \emph{snapshot image} only averages over source noise, an \emph{average image} additionally averages over the diffractive scintillation, and an \emph{ensemble-average} image averages over both diffractive and refractive scintillation. 

An extended source quenches the scintillation. Sources with angular size much larger than $r_0/D$ quench the diffractive scintillation, while sources with angular size much larger than $r_{\rm R}/D$ quench the refractive scintillation. For extremely compact sources, such as pulsars, there are now a number of promising techniques to determine the coherent scattering response by utilizing the point-like nature of the pulsed signal and the rich information in the diffractive scintillation pattern \citep[e.g.,][]{Walker_2008,Brisken_2010}. However, our focus is observations of strongly-scattered sources that can resolve the intrinsic angular structure. We will therefore assume that the diffractive scintillation is quenched (the refractive scintillation may also be partially quenched). Consequently, a ``single-epoch'' image corresponds to the average image in the language of Narayan and Goodman, or a ``speckle image'' in the language of optical scattering.

\citet{Blandford_Narayan_1985} argued that diffractive and refractive scattering effects on images can be treated separately. Because diffractive effects are dominated on small scales (${\sim}r_0$), they can be approximated by their ensemble-average effect: blurring of the unscattered image with a kernel $G(\mathbf{r})$ (the seeing disk). This kernel is most naturally represented in the Fourier-conjugate visibility domain, $\tilde{G}(\mathbf{b}) = e^{-\frac{1}{2}D_{\phi}(\mathbf{b}/(1+M))}$, where $\mathbf{b}$ corresponds to the physical length of an interferometric baseline. Refractive effects can be approximated in a geometrical optics framework -- gradients of the large-scale, refractive modes of the phase screen will steer and focus the ensemble-average image. The Appendix of \citet{Johnson_Narayan_2016} provides formal justification for this approach via the Fresnel diffraction integral.  The single-epoch scattered image $I_{\rm a}(\mathbf{r})$ is then related to the unscattered image $I_{\rm src}(\mathbf{r})$ via \citep[][Eq.~9,~10]{Johnson_Narayan_2016}
\begin{align}
\label{eq::Geometric_Optics}
I_{\rm a}(\mathbf{r}) &\approx I_{\rm ea}\left(\mathbf{r} + r_{\rm F}^2 \nabla \phi_{\rm r}(\mathbf{r}) \right)\\
\nonumber &\approx I_{\rm ea}(\mathbf{r}) + r_{\rm F}^2 \left[\nabla \phi_{\rm r}(\mathbf{r})\right] \cdot \left[ \nabla I_{\rm ea}(\mathbf{r}) \right]\\
\nonumber &= I_{\rm src}(\mathbf{r}) \ast G(\mathbf{r}) + r_{\rm F}^2 \left[\nabla \phi_{\rm r}(\mathbf{r})\right] \cdot \left[ \nabla \left( I_{\rm src}(\mathbf{r}) \ast G(\mathbf{r}) \right) \right].
\end{align}
In these expressions, $\nabla$ denotes a two-dimensional, transverse gradient on the phase screen, and we have written the ensemble-average image as $I_{\rm ea}(\mathbf{r}) = I_{\rm src}(\mathbf{r}) \ast G(\mathbf{r})$, where $\ast$ is a spatial convolution. For each image, $\mathbf{r}$ is a transverse coordinate at the distance of the scattering screen $D$ (not the distance of the source, $D+R$), so the corresponding angular scales are $\boldsymbol{\theta} = \mathbf{r}/D$.  For simplicity, throughout the remainder of this paper we will use $\phi(\mathbf{r})$ to denote the refractive phase screen $\phi_{\rm r}(\mathbf{r})$. 

Figure~\ref{fig::Imaging_Overview} summarizes the relationships between the unscattered, average, and ensemble-average images after decoupling the diffractive and refractive scattering.

\subsection{Refractive Substructure}

As Figure~\ref{fig::Imaging_Overview} shows, refractive phase gradients ``shuffle'' image brightness, producing substructure in scattered images \citep{NarayanGoodman89,GoodmanNarayan89}. Implications of refractive substructure for VLBI of extended sources have been calculated and discussed in detail by \citet{Johnson_Gwinn_2015}, \citet{Johnson_2016}, and \citet{Johnson_Narayan_2016}. In particular, refractive substructure introduces a new source of noise for interferometric observables. However, this ``refractive noise'' is significantly different from thermal noise. Refractive noise is wideband and persistent, with a decorrelation bandwidth of order unity and typical decorrelation timescale of days to weeks. Refractive noise is also correlated among different baselines and is sensitive to the intrinsic source structure. Thus, modeling the refractive noise \emph{requires} a source model, and some signatures of the refractive noise, such as closure phase jitter from scattering, can be very sensitive to the source model \citep[see, e.g., Fig.~4 of][]{Johnson_Narayan_2016}. Most importantly, refractive noise falls slowly with increasing baseline length (the rms amplitude falls roughly as $\left| \mathbf{b} \right|^{-5/6}$), so it can provide significant power on long baselines, \emph{even on baselines that would have entirely resolved the unscattered source}. Consequently, refractive substructure can profoundly influence VLBI imaging and is increasingly important at higher angular resolution.

\begin{figure}[t]
\centering
\includegraphics[width=0.45\textwidth]{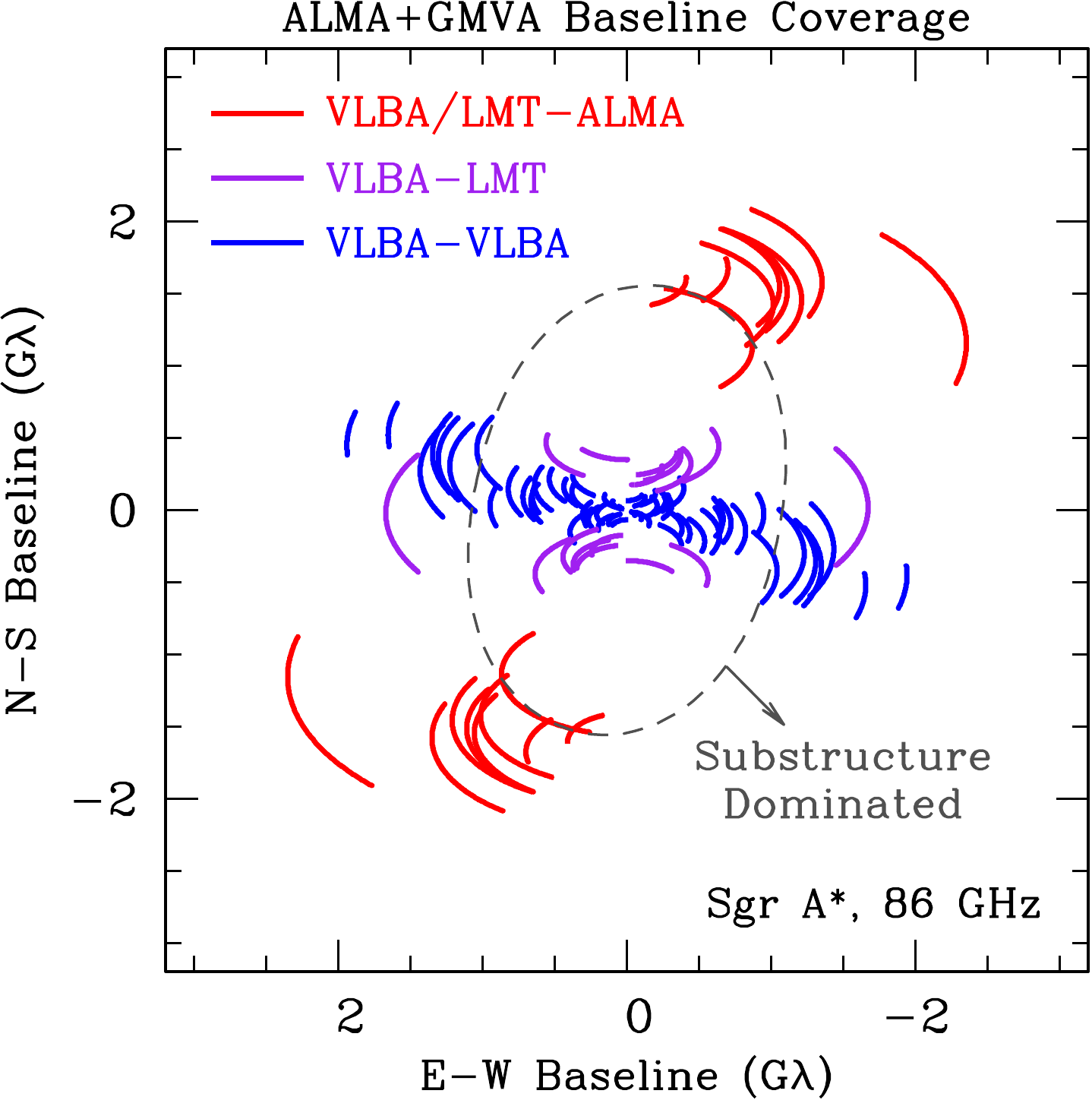} 
\caption
{ 
Baseline coverage for \sgra\ with the VLBA, LMT, GBT, and ALMA at 3.5\,mm wavelength. ALMA brings many long, sensitive baselines but detections on these baselines will be strongly affected by refractive substructure. The dashed ellipse denotes the transition between baselines that are dominated by the ensemble-average visibility (within the ellipse) to baselines that are dominated by the random fluctuations from refractive substructure (outside the ellipse). This calculation assumed an unscattered source that is a circular Gaussian with FWHM of 130\,$\mu$as, as is consistent with current observations \citep{Lu_2011,Ortiz_2016}. However, the size of the transition ellipse is not especially sensitive to the size of the unscattered source because a smaller/larger intrinsic source will raise/lower both the ensemble-average visibility and the refractive noise on long baselines.
}
\label{fig::SgrA_uv}
\end{figure}

Because detecting refractive substructure requires VLBI with both high sensitivity and extremely high angular resolution, unambiguous signatures of refractive substructure have only recently been identified. For \sgra\ at 1.3\,cm, the detection of substructure on long baselines was enabled by using the GBT with the VLBA recording 512\,MHz of bandwidth \citep{Gwinn_2014}. At 3.5\,mm, recently detected non-zero closure phases are likely associated with substructure and were only measured when the LMT was used in concert with the VLBA \citep{Ortiz_2016,Brinkerink_2016}. At 1.3\,mm, the closure phase on a triangle of baselines joining California, Arizona, and Hawaii has shown persistent sign over four years, demonstrating that refractive noise cannot produce the observed non-zero signal \citep{Fish_2016}; however, an analysis based on radiatively inefficient accretion flow models found that the observed closure phase variability may be dominated by refractive noise \citep{Broderick_2016}. The addition of the Atacama Large Millimeter/submillimeter Array (ALMA) to VLBI experiments in 2017 \citep{Fish_2013} will sharply increase the sensitivity at 3.5 and 1.3\,mm and should lead to many baselines with strong detections that are dominated by the stochastic signal from refractive substructure (see Figure~\ref{fig::SgrA_uv}). Consequently, scattering will strongly affect VLBI imaging of \sgra\ with ALMA, and a suitable mitigation framework is essential to derive robust conclusions from these images.

\section{Stochastic Optics}
\label{sec::Mitigation_Framework}

We now develop a new scattering mitigation framework for VLBI imaging. We begin by summarizing general considerations for mitigation strategies, then we describe the new framework and details about how it can be implemented. We will conclude this section with example reconstructions using this framework on simulated data. 

\subsection{General Considerations}

The goal of our mitigation framework is two-fold: to partially invert the ``blurring'' from diffractive scattering, and to identify and minimize spurious image features caused by refractive scattering. Although the first goal is identical for radio and optical scattering (essentially, developing a technique that brings the Strehl ratio closer to unity), the second is specific to the radio case because of the vastly different timescales involved. For optical mitigation, the coherence timescale is milliseconds so the observations can collect a representative ensemble of independent scattered images. However, even at the highest current resolutions of VLBI, the relevant coherence timescale for resolved images is still at least $\left(\frac{\theta_{\rm uv}}{20~\mu{\rm as}} \right) \left( \frac{D}{1~{\rm kpc}} \right) \left( \frac{V_{\perp}}{50~{\rm km/s}} \right)^{-1} \times 17~{\rm hours}$. Consequently, a VLBI observation will typically only sample a single realization of the scattering, so it is necessary but challenging to distinguish intrinsic image features from scattering-induced features. 

Further difficulties in developing a successful scattering mitigation framework for radio observations are apparent from Eq.~\ref{eq::Geometric_Optics}. The imprint of scattering is stochastic, scale-dependent, and variable across an image. Unlike blurring of the ensemble-average image, single-epoch scattering cannot be unambiguously identified or removed (even with a perfect reconstruction of the scattered image), so mitigation frameworks must distinguish intrinsic structure from scattering using a statistical approach. Also, the scattering can introduce large variations in interferometric visibilities that are strongly correlated among groups of visibilities with similar baselines and frequencies. In this regard, the imprint of scattering is similar to amplitude and phase calibration errors, which can likewise be correlated over long periods of time. However, the scattering is \underline{not} a station-dependent effect, so it cannot be inverted by a suitable choice of complex, time-variable station gains. And because the imprint of refractive noise depends on the unscattered image, the refractive noise from scattering cannot be assessed independently of the imaging process (e.g., by simply adding additional terms to the error budget of measured visibilities). 

Despite these difficulties, the scattering still exhibits many convenient properties. For example, the scattering is not significantly birefringent and it has a deterministic scaling with frequency ($\phi \propto \lambda$ and $r_{\rm F}^2 \nabla \phi(\mathbf{r}) \propto \lambda^2$). After averaging over time to approach the ensemble-average regime, the scattering reduces to a convolution with a deterministic kernel. And, in some cases, the statistical properties of the scattering, such its time-averaged power spectrum, $Q(\mathbf{q})$, may be well known from previous measurements. 

\subsection{Mitigation Framework}

We now develop a strategy that utilizes known properties of scattering while accounting for unknown, stochastic perturbations. Our general procedure is analogous to adaptive optics, but it does not mitigate the scattering through instrumentation (either physically or digitally, through compensating station-based phases). Because of this fundamental difference and because of the dependence on specified statistical properties of the scattering, we will refer to this new mitigation framework as ``stochastic'' optics.  

To proceed, we use the approximate representation of the scattered image given in Eq.~\ref{eq::Geometric_Optics}. For the purposes of scattering mitigation, the most important property of Eq.~\ref{eq::Geometric_Optics} is that the scattered image is described entirely in terms of the ensemble-average image and its large-scale, refractive perturbations (see Figure~\ref{fig::Imaging_Overview}). For image reconstructions, it is not necessary to expand beyond $I_{\rm a}(\mathbf{r}) \approx I_{\rm ea}\left(\mathbf{r} + r_{\rm F}^2 \nabla \phi(\mathbf{r}) \right)$ in Eq.~\ref{eq::Geometric_Optics}. In fact, the linear expansion in the second line of Eq.~\ref{eq::Geometric_Optics} is not constrained to be positive and so typically produces a faint halo of negative flux around bright regions in the image, so it may be advantageous to keep the first form. However, for simplicity, we will employ the linear approximation for the remaining discussion and results. 

We assume that the unscattered image $I_{\rm src}(\mathbf{r})$ and the large-scale phase screen $\phi(\mathbf{r})$ are unknown but that the diffractive kernel $G(\mathbf{r})$ and the time-averaged scattering power spectrum $Q(\mathbf{q})$ are known. The imaging problem is then to use a set of measured visibilities sampled from the scattered image $I_{\rm a}(\mathbf{r})$ to simultaneously estimate the pair of images $I_{\rm src}(\mathbf{r})$ and $\phi(\mathbf{r})$.

\subsection{Discrete Representation of the Scattering Screen}
\label{sec::DiscreteScatteringScreen}

For our mitigation framework, $\phi(\mathbf{r})$ is most naturally represented in the Fourier domain. The Fourier components $\tilde{\phi}(\mathbf{q}) = \int d^2\mathbf{r}\, \phi(\mathbf{r}) e^{-i \mathbf{q} \cdot \mathbf{r}}$ are uncorrelated, complex Gaussian random variables. The time-averaged power spectrum is given by 
\begin{align}
\left \langle \left|\tilde{\phi}(\mathbf{q}) \right|^2 \right \rangle= \lambdabar^{2} A_{\phi} Q(\mathbf{q}), 
\end{align}
where $A_{\phi}$ is the screen area over which the Fourier transform is computed \citep[see, e.g.,][]{Blandford_Narayan_1985,Goodman_1987}. 

We parametrize the scattering phase screen using an $N \times N$ grid of Fourier coefficients $\tilde{\phi}_{s,t}$:
\begin{align}
\label{eq::phi_epsilon}
 \phi_{\ell,m} &= \frac{1}{F^2} \sum_{s,t=0}^{N-1} \tilde{\phi}_{s,t} e^{2\pi i \left( \ell s + m t \right)/N}\\
\nonumber      &\equiv \frac{\lambdabar}{F}\sum_{s,t=0}^{N-1} \sqrt{Q(s,t)} \times \epsilon_{s,t} e^{2\pi i \left(  \ell s + m t  \right)/N}\!.
\end{align}
In this expression, $F$ is the image field of view expressed as a transverse length on the scattering screen (i.e., $F^2 = A_{\phi}$). The image resolution is then $F/N$, and the spectral resolution is $2\pi/F$. We have further simplified the representation of the conjugate screen phase by introducing a set of standardized, complex Gaussian random variables $\epsilon_{s,t} \equiv \tilde{\phi}_{s,t}/\sqrt{Q(s,t)}$ (i.e., $\langle \left| \epsilon_{s,t} \right|^2 \rangle = 1$). To ensure that $\phi_{\ell,m} \in \mathbb{R}$, we require that $\epsilon_{s,t} = \epsilon_{-s,-t}^\ast$ (where negative indices are wrapped: $-s \rightarrow N-s$ for $s>0$). And because a constant offset of the phase does not affect the scattered image, we set the mean phase to be zero: $\epsilon_{0,0} = 0$. The unknown refractive scattering screen $\phi(\mathbf{r})$ is then parameterized by $N_\phi \equiv (N^2-1)/2$ independent complex elements of ${\epsilon}_{s,t}$. 

Once $\epsilon_{s,t}$ has been specified for a particular field of view, the corresponding scattering can be computed for that field of view for any desired observing wavelength. Because $\phi_{\ell, m} \propto \lambda$ (from the cold plasma dispersion law) and $r_{\rm F} \propto \sqrt{\lambda}$, the refractive steering angles have a deterministic wavelength dependence: $r_{\rm F}^2 \nabla \phi(\mathbf{r}) \propto \lambda^2$. Note that these scalings are independent of the power-law parameter $\alpha$ of the phase and density fluctuations, which instead governs the shape of the diffractive blurring kernel and the relative power in phase fluctuations at different wavenumbers.

\subsection{Imaging Procedure}

Image synthesis via stochastic optics simultaneously estimates the pair of images $I_{\rm src}(\mathbf{r})$ and $\phi(\mathbf{r})$ over a prescribed field of view. These are parameterized by the two discrete arrays $\mathbf{I}_{\rm src}$ and $\boldsymbol{\epsilon}$. As with all VLBI imaging, the problem of image reconstruction is ill-posed and must be regularized. Many regularizers are commonly used for imaging \citep[see, e.g.,][]{Thiebaut_2013,Bouman_2016}; for specificity, we will use the standard maximum entropy method \citep[MEM; see][]{Narayan_Nityananda_1986} to regularize the unscattered image $\mathbf{I}_{\rm src}$. 

The conventional imaging approach (which has no scattering mitigation) is to find the image $\mathbf{I}_{\rm src}$ that maximizes the objective function
\begin{align}
 J = S(\mathbf{I}_{\rm src}; \mathbf{B}) - \alpha_{\rm V} \chi^2_{\rm V}.
\end{align}
In this expression, $S(\mathbf{I}_{\rm src}; \mathbf{B})$ denotes the entropy function for the unscattered image, and $\chi^2_{\rm V}$ represents a chi-squared for whatever data products are used as part of the imaging. These data products can include complex visibilities that have been self-calibrated, the bispectrum, or the set of all closure amplitudes and phases, for instance. We have also included a bias image, $\mathbf{B}$, which can optionally be used to refine the imaging according to a priori expectations for the image extent or morphology. $\alpha_{\rm V}$ is a ``hyperparameter'' that controls the relative weighting of the entropy and data terms. It can be adjusted manually or automatically to yield the expected $\chi^2_{\rm V}$ for a satisfactory image \citep[e.g.,][]{Cornwell_Evans_1985}. Scattering mitigation techniques based on deconvolution or ``deblurring'' can also use this framework by modifying the data chi-squared term, instead using visibilities and noise that have been appropriately scaled (we discuss these techniques in detail in \S\ref{sec::Deblurring}).

This imaging framework can be motivated through a Bayesian approach wherein the objective function $J$ is associated with the log posterior probability of the reconstructed image. With this perspective, the extension to include scattering is straightforward. Because the scattering is defined by an array $\boldsymbol{\epsilon}$ of independent, standardized, complex Gaussian random variables, the log likelihood of $\boldsymbol{\epsilon}$ is simply $\ln \mathcal{L} = -\left( \chi^2_{\phi} + N_{\phi}\ln \pi \right)$, where $\chi^2_{\phi} \equiv \sum \left|\epsilon_{s,t}\right|^2$ and the sum ranges over only the $N_{\phi}$ independent, non-zero elements $\epsilon_{s,t}$. 
Thus, to image both the source and the scattering, one can maximize the objective function
\begin{align}
\label{eq::J_scattering}
 J = S(\mathbf{I}_{\rm src}; \mathbf{B}) - \alpha_{\rm V} \chi^2_{\rm V} - \alpha_{\phi}\chi^2_{\phi}.
\end{align}
Note that $\chi^2_{\rm V}$ now must be calculated with respect to the scattered image $\mathbf{I}_{\rm a}$, which is a function of $\mathbf{I}_{\rm src}$, $\boldsymbol{\epsilon}$, and the ensemble-average scattering kernel $G(\mathbf{r})$ and power spectrum $Q(\mathbf{q})$ (see Eq.~\ref{eq::Geometric_Optics} and Eq.~\ref{eq::phi_epsilon}). The addition of $\chi_\phi^2$ to the objective function ensures that the power in the estimated screen phases $\phi_{\ell,m}$ is compatible with $Q(\mathbf{q})$. 
We have included an additional hyperparameter $\alpha_{\phi}$ to control the relative strength of this regularization. 

A convenient property of Eq.~\ref{eq::J_scattering} is that high-frequency elements of $\epsilon_{s,t}$ that correspond to finer angular scales than the array resolution (and are therefore unconstrained by the visibility data) will approach 0 upon successful convergence to a maximum in $J$. Because of this property, the resulting normalized value $N_{\phi}^{-1} \chi^2_{\phi}$ may be much less than unity after successful convergence, depending on how the pixel resolution compares with the nominal array resolution. To avoid this result and to reduce the complexity of the image synthesis, one could limit the number of Fourier components in $\boldsymbol{\epsilon}$ to model only those angular scales that are accessible to the participating baselines. However, in our imaging examples later, we do not limit the number of Fourier components in $\boldsymbol{\epsilon}$; we instead assess suitable convergence by examining the maximum value of the reconstructed $|\epsilon_{s,t}|$. Because the true values $|\epsilon_{s,t}|$ are each drawn from a Rayleigh distribution, the maximum of $n$ such values is highly peaked and insensitive to $n$; specifically, the maximum is an approximately Gaussian random variable with mean $\sqrt{\ln n}$ and standard deviation $0.5/\sqrt{\ln n}$ \citep{TMS}. The applicable $n$ is roughly the number of beams per image, which is characteristically between $10^2$ and $50^2$ for \sgra\ at millimeter wavelengths (see the examples in \ref{sec::Example_Reconstructions}). Thus, as a rough rule of thumb, we expect ${\rm max}\left(|\epsilon_{s,t}|\right) \sim 2.5$, and we adjust the hyperparameter $\alpha_{\phi}$ accordingly.  

Our particular choice of objective function could easily be adapted to whatever regularization for the unscattered image is desired, although there is not an obvious extension to different classes of imaging algorithms such as CLEAN \citep{Hogbom_1974}. Eq.~\ref{eq::J_scattering} can also be generalized to image multiple frequencies or polarizations simultaneously \citep[see, e.g.,][]{Chael_2016}, and it can incorporate temporal evolution of the scattering using the standard ``frozen-screen'' approximation \citep{Taylor_1938}.

\afterpage{
\begin{figure*}[h!]
\centering
\includegraphics[width=1.0\textwidth]{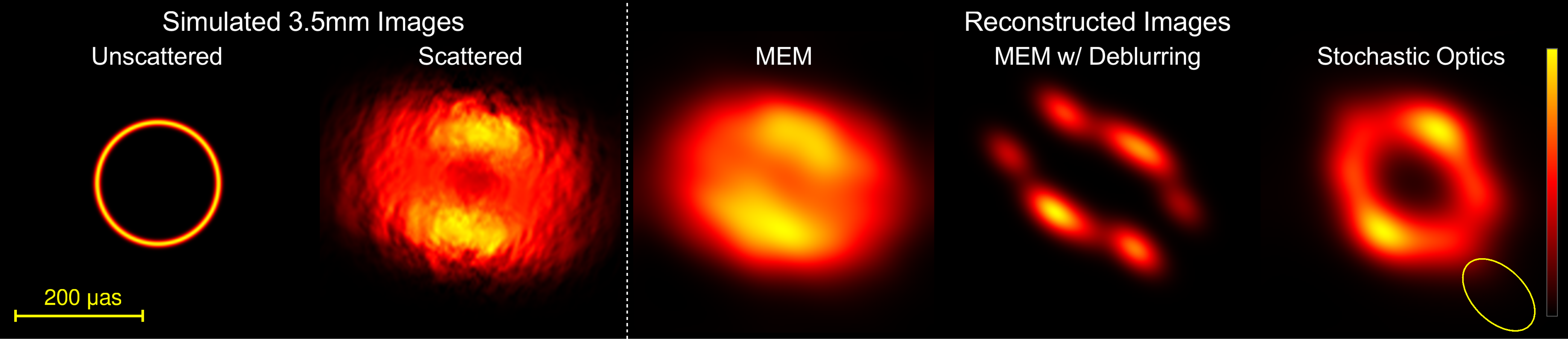}
\caption
{ 
The leftmost panels show the ring model from Figure~\ref{fig::Imaging_Overview} before and after scattering at 3.5\,mm. The remaining three panels compare three image reconstruction strategies: (center) MEM with no scattering mitigation, (center right) MEM with ``deblurred'' visibilities \citep{Fish_2014}, and (right) stochastic optics. Following the methodology of \citet{Chael_2016}, each of the reconstructions is convolved with half of the CLEAN beam (the full beam is indicated in the lower right). The color scale of each image is linear and is indicated on the right.
}
\label{fig::Stochastic_Optics_Ring}
\end{figure*}
\begin{figure*}[t]
\centering
\includegraphics[width=1.0\textwidth]{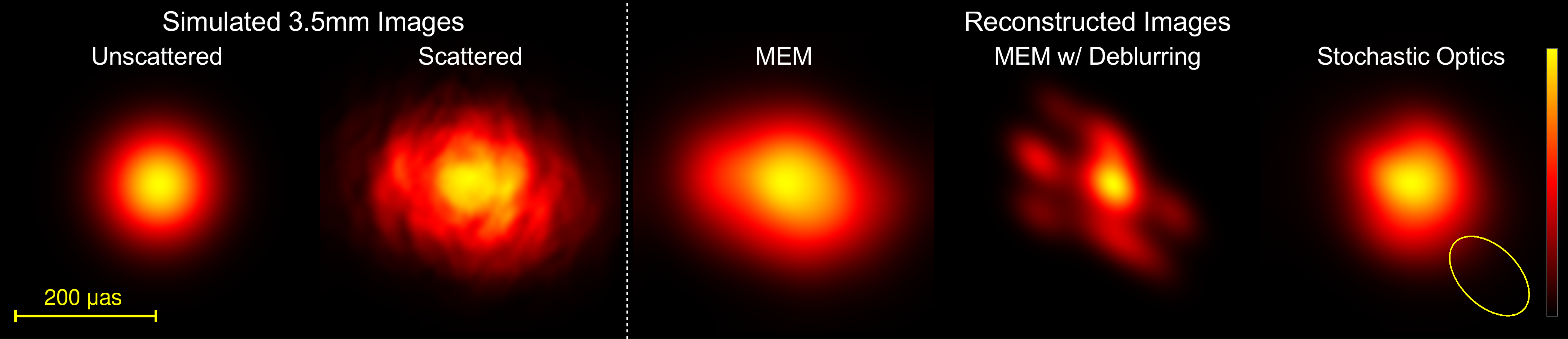}\\
\includegraphics[width=1.0\textwidth]{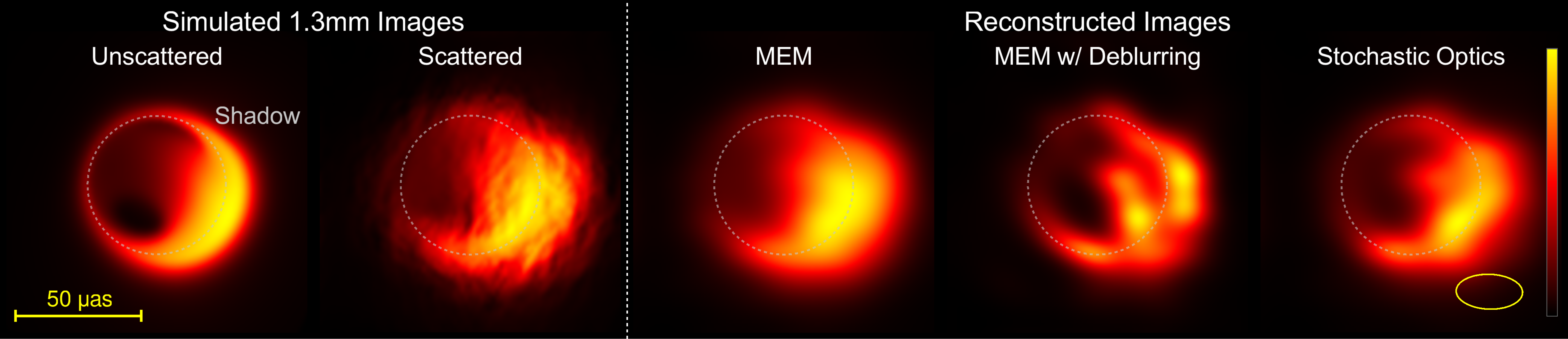}
\caption
{ 
The leftmost panels show simulated images of \sgra\ at 3.5\,mm (top) and at 1.3\,mm (bottom) before and after scattering. The unscattered 3.5\,mm image is a $130\,\mu{\rm as}$ circular Gaussian with a total flux density of 3.4~Jy, which roughly matches the observed size and flux density of \sgra\ at this wavelength \citep{Lu_2011,Ortiz_2016}, while the unscattered 1.3\,mm image is a ray-traced semi-analytic accretion flow model for \sgra\ with a total flux density of 2.4~Jy \citep{Broderick_2016}. For reference, the dotted circle indicates the circular ring corresponding to the black hole ``shadow'' for \sgra\ \citep{Bardeen_1973,Luminet_1979,Falcke_2000}. The remaining panels compare three image reconstruction strategies: (center) MEM with no scattering mitigation, (center right) MEM with ``deblurred'' visibilities, and (right) stochastic optics. As in Figure~\ref{fig::Stochastic_Optics_Ring}, the color scale of each image is linear, and each of the reconstructions is convolved with half of the CLEAN beam (the full beam is indicated in the lower right). For these examples, the reconstructions at the two frequencies were performed independently.
}
\label{fig::Imaging_Comparisons}
\end{figure*}
\begin{figure*}[t]
\centering
\includegraphics[height=4.2cm,trim={0.3cm 0.3cm 0.1cm 0.3cm}, clip]{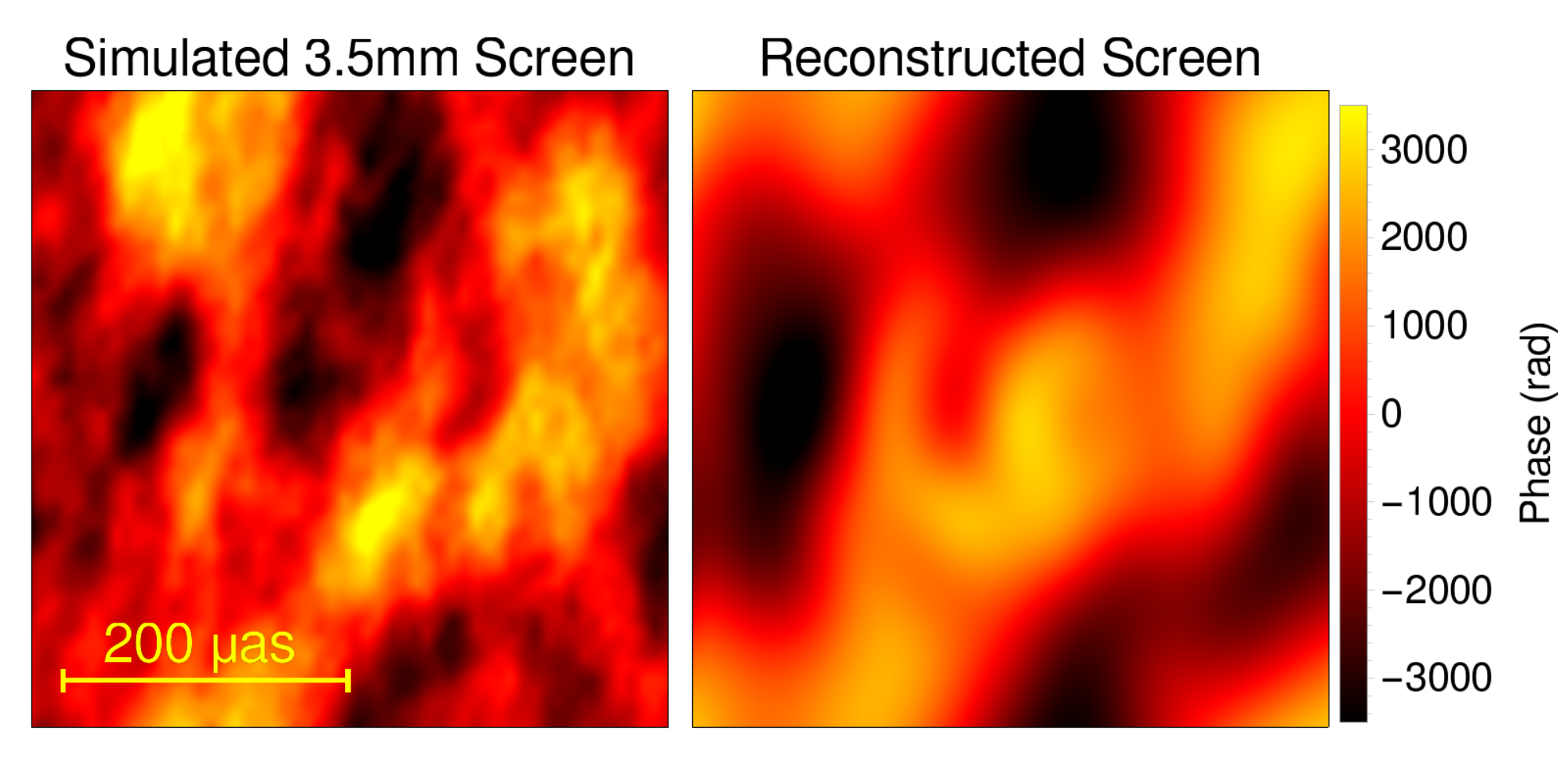}\hfill
\includegraphics[height=4.2cm,trim={0.3cm 0.3cm 0.1cm 0.3cm}, clip]{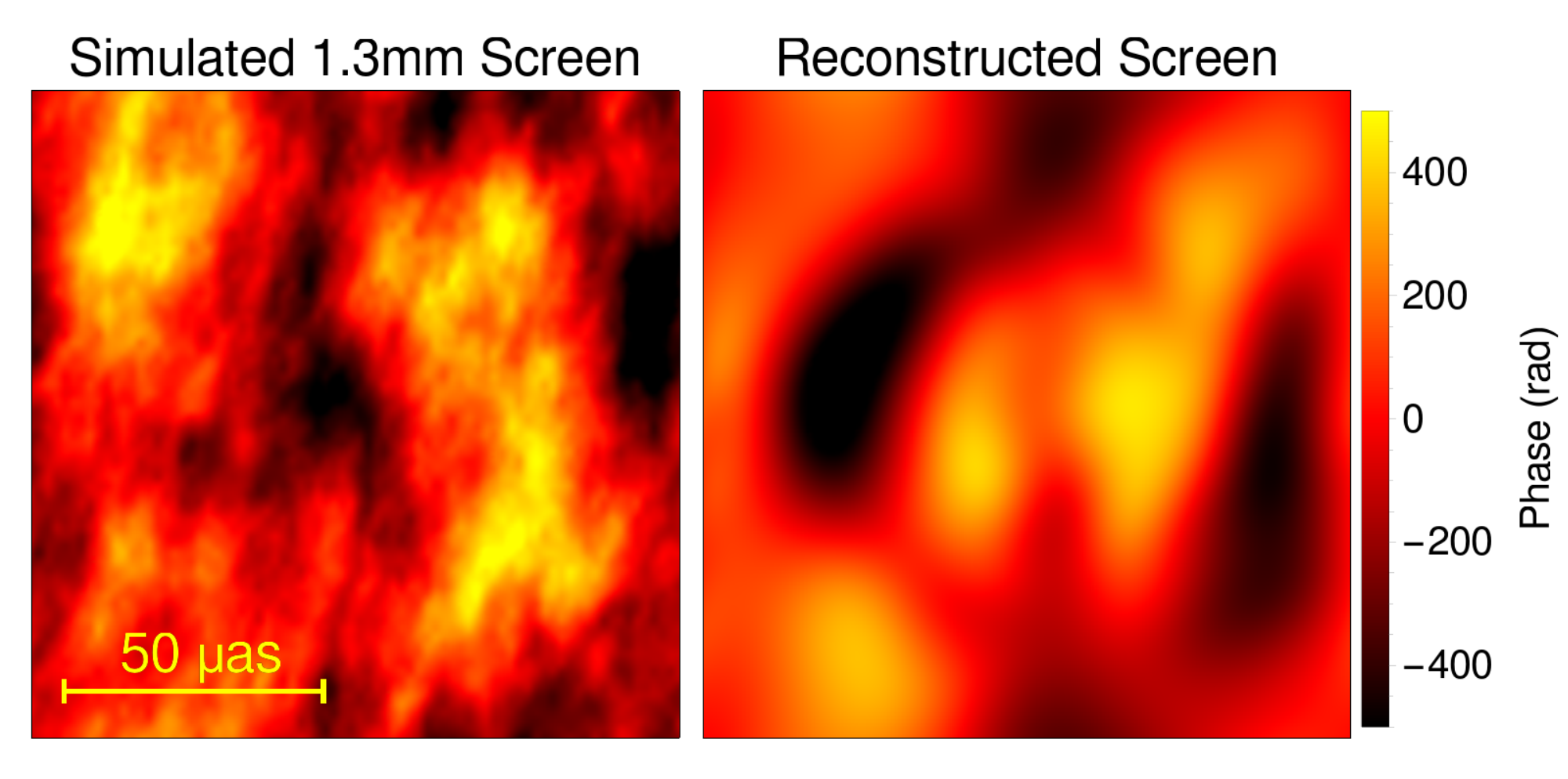}
\caption
{ 
Simulated and reconstructed phase screens for the examples of Figure~\ref{fig::Imaging_Comparisons}. For these comparisons, in addition to setting $\epsilon_{0,0} = 0$ we also set $\epsilon_{\pm 1,0} = \epsilon_{0,\pm 1} = 0$ to eliminate any overall image shift from scattering, which is degenerate with the choice of centroid for the unscattered image in these examples. The reconstructed phase screen at 3.5\,mm is more accurate than at 1.3\,mm because there are more baselines that are dominated by refractive substructure (see Figure~\ref{fig::SgrA_uv}). In each case, the reconstructed phase screen is only constrained near regions of the image with non-zero flux density in the ensemble-average image.
}
\label{fig::Reconstructed_Phase_Screens}
\end{figure*}
\begin{figure*}[t]
\centering
\includegraphics[width=1.0\textwidth]{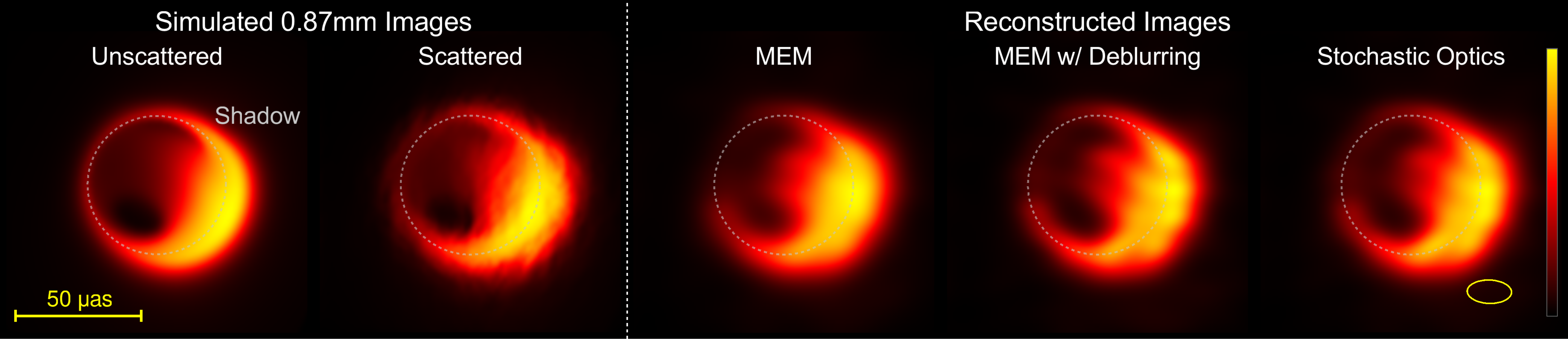}
\caption
{ 
Same as the bottom panels of Figure~\ref{fig::Imaging_Comparisons}, except that the scattering and observations correspond to 345~GHz instead of 230~GHz. In this case, the three different reconstructions all give similar results, and the differences between MEM and stochastic optics are insignificant. 
}
\label{fig::345_GHz_Example}
\end{figure*}
}

\subsection{Example Reconstructions}
\label{sec::Example_Reconstructions}

We now provide example image reconstructions using stochastic optics. We will begin with an example that is illustrative but physically implausible before considering more realistic unscattered images. In all cases, we will use the measured scattering properties of \sgra. Specifically, we assume that the diffractive kernel is an elliptical Gaussian with FWHM of $\left\{ 1.309,\, 0.64\right\}\times \lambda_{\rm cm}^2~{\rm mas}$ and with the major axis position angle $78^\circ$ east of north \citep[][]{Bower_2006}. We assume that the large-scale power spectrum follows a Kolmogorov scaling, $\alpha=5/3$, which now has tentative support for \sgra\ \citep{Gwinn_2014}. We further assume that the large-scale power has the same anisotropy as the diffractive kernel, and so we use the form \citep[][Eq.~6]{Johnson_Narayan_2016}
\begin{align}
Q(\mathbf{q}) &= 2^{\alpha} \pi \alpha \frac{\Gamma\left(1+\alpha/2\right)}{\Gamma\left(1-\alpha/2\right)} \lambdabar^{-2} \left( r_{0,x} r_{0,y} \right)^{-\alpha/2}\\
\nonumber & \qquad \times \left[ \left(\frac{r_{0,x}}{r_{0,y}}\right) q_{x}^2 + \left(\frac{r_{0,y}}{r_{0,x}}\right) q_{y}^2 \right]^{-(1 + \alpha/2)}\!,
\end{align}
where $\{ x, y\}$ are the coordinates aligned with the major and minor axes of the diffractive kernel. We adopt a scattering screen located at a fractional distance of 0.30 to the source ($M=0.43$) \citep{Bower_2014}. This gives $r_{0,x} = 412 \lambda^{-1}\,{\rm km}$ and $r_{0,y} = 844 \lambda^{-1}\,{\rm km}$.

We implemented the scattering and stochastic optics in Python by extending a library that was originally developed for polarimetric synthesis imaging \citep[\url{https://github.com/achael/eht-imaging;}][]{Chael_2016}. Because it is straightforward to estimate analytic gradients of $J$, even when the scattering terms are included (Eq.~\ref{eq::J_scattering}), the imaging does not require significant computing resources. For the images in this section, we used reconstructions with $55\times 55~{\rm pixels}$, which each took only a few minutes to generate on a personal computer. 

For each image, we generated synthetic data according to projected array configurations and performance in 2017 (see Table~\ref{tab::Arrays}). Although system equivalent flux densities (SEFDs) are higher at 1.3\,mm than at 3.5\,mm, the observing sensitivities are comparable because of the wider recorded bandwidth with current 1.3\,mm systems (4~GHz vs.\ 512~MHz). To simplify the imaging implementation and comparisons, we generated data with expected thermal noise but no systematic uncertainties. In the future, we will develop extensions that account for systematic uncertainties, both through the use of data products that eliminate these uncertainties (such as the bispectrum, closure phases, and closure amplitudes) and through iterative self-calibration. These extensions will be a critical precursor to using stochastic optics on actual VLBI data, but our primary focus here is simply to evaluate whether or not stochastic optics can reliably decouple the intrinsic and scattering structure and to assess its performance relative to existing mitigation strategies. We used the standard entropy function $S(\mathbf{I}) = -\sum_{\ell, m} I_{\ell, m} \ln \left( I_{\ell, m} \right)$ \citep{Frieden_1972,Gull_Daniell_1978}. 

For each image reconstruction, we began with an ordinary MEM reconstruction. We followed the procedure shown in the upper half of Figure~1 from \citet{Chael_2016}, repeatedly blurring the converged image and re-imaging to help ensure that the final solution was a global maximum for $J$. We then used the resultant image (an approximation of the average image) as an initial guess and bias image for $\mathbf{I}_{\rm src}$ and then imaged the data using the stochastic optics framework. We again repeatedly blurred and re-imaged the data to ensure global convergence. We also re-initialized the phase screen to be uniformly zero for each of the repeated imaging steps. With this procedure, all of the images converged successfully without user intervention or supervision.

\begin{deluxetable}{lr}
\tablewidth{\columnwidth}
\tablecaption{Array Parameters for Example Reconstructions.}
\tablehead{
\colhead{ {\bf Site} } & \colhead{ {\bf SEFD (Jy)} } }
\startdata
\sidehead{\bf 3.5\,mm (GMVA)}
VLBA (${\times}8$) & 2500 \\
Green Bank Telescope (GBT) & 137 \\
Large Millimeter Telescope (LMT) & 1714 \\
ALMA & 75 \\
\tableline
\sidehead{\bf 1.3\,mm (EHT)}
Submillimeter Array (SMA) & 4000\\
Submillimeter Telescope (SMT) & 1100\\
LMT & 1400\\
ALMA & 100\\
IRAM 30m & 1400\\
NOEMA single dish & 5200\\
South Pole Telescope (SPT) & 9000
\enddata
\label{tab::Arrays}
\tablecomments{Expected VLBI arrays in 2017. The VLBA at 3.5\,mm includes Fort Davis, Pie Town, Los Alamos, Kitt Peak, Mauna Kea, Brewster, North Liberty, and Owens Valley (each VLBA site has an estimated SEFD of 2500\,Jy). At 3.5\,mm, the expected bandwidth is 512~MHz; at 1.3\,mm, the expected bandwidth is 4~GHz. Array parameters at 3.5\,mm were taken from \url{http://www3.mpifr-bonn.mpg.de/div/vlbi/globalmm/}. Parameters at 1.3\,mm were taken from \url{http://www.eventhorizontelescope.org/proposal.html}.}
\end{deluxetable}

\begin{figure*}[t]
\centering
\includegraphics[width=1.0\textwidth]{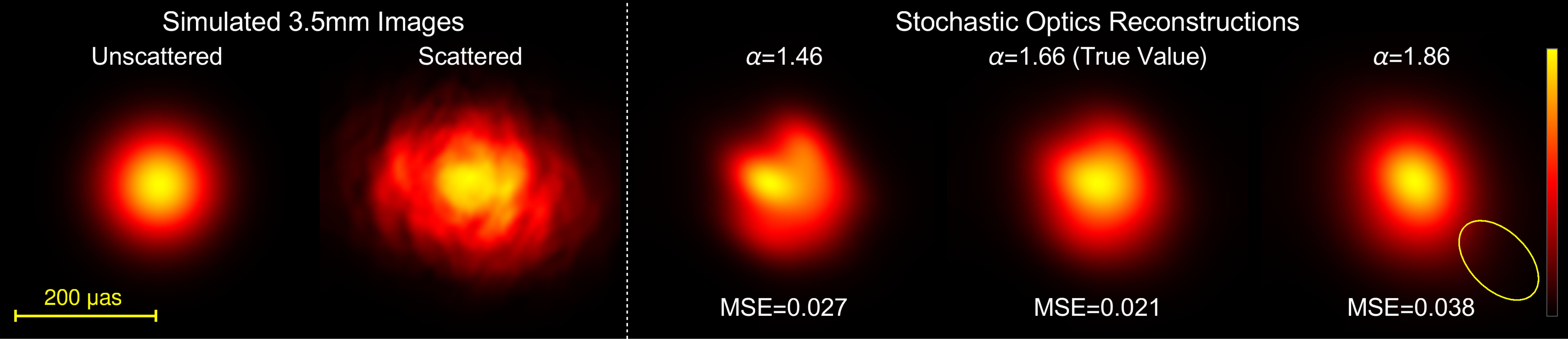}
\includegraphics[width=1.0\textwidth]{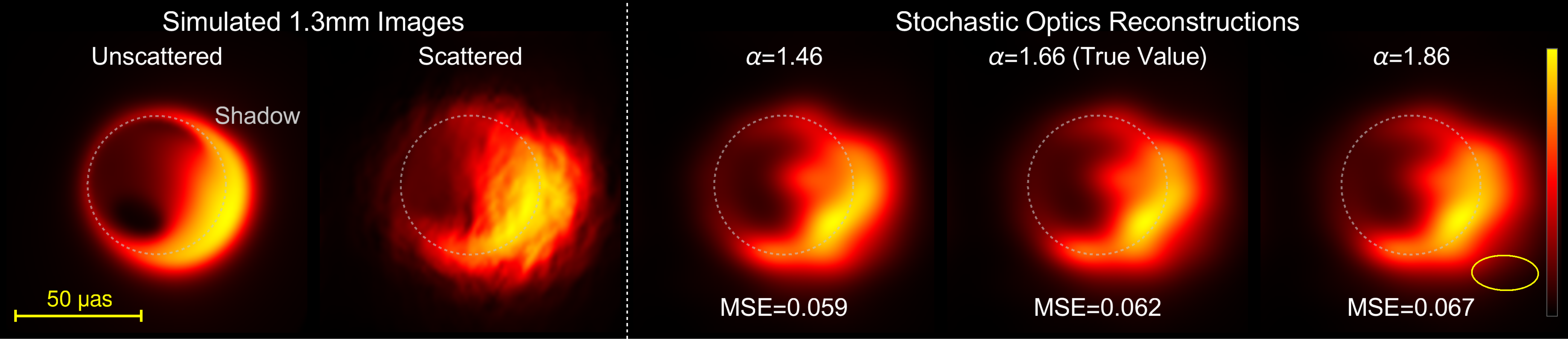}
\caption
{ 
Comparison of stochastic optics reconstructions when the assumed value of $\alpha$ is varied (the simulated images have $\alpha=5/3$). Reconstructions that assume $\alpha=1.46$ slightly underpredict the level of substructure, resulting in synthesized images with more noticeable remaining artifacts from refractive scattering; reconstructions that assume $\alpha=1.86$ slightly overpredict the level of substructure, resulting in synthesized images that are overly smooth. However, these changes are relatively mild, especially at 1.3\,mm (e.g., the reconstruction with $\alpha=1.46$ even has a slightly lower MSE than the reconstruction with the correct value of $\alpha$). 
}
\label{fig::Varying_Alpha}
\end{figure*}

Figure~\ref{fig::Stochastic_Optics_Ring} shows the results of stochastic optics when applied to the scattered ring shown in Figure~\ref{fig::Imaging_Overview}. Although this model is not physically motivated, we set the total flux density of the ring to be 3.4~Jy, matching \sgra\ at 3.5\,mm \citep{Lu_2011}, so that the thermal noise in our synthetic data was roughly consistent with observations. Figure~\ref{fig::Stochastic_Optics_Ring} compares reconstructions with MEM, with MEM after ``deblurring'' the measured visibilities (see \S\ref{sec::Deblurring}), and with stochastic optics. Stochastic optics easily outperforms the alternatives, and the remaining artifacts --- i.e., bright antipodal lobes --- are primarily a result of the restoring beam, which produces brightest regions when the major axis of the beam is tangent to the ring (an anisotropic scattering kernel has the same effect; see Figure~\ref{fig::Imaging_Overview}). Note that stochastic optics successfully removes much of the diffractive blurring and also corrects the large-scale image distortions that are apparent in the deblurred reconstruction. 

Figure~\ref{fig::Imaging_Comparisons} compares reconstructions for more realistic unscattered images at 1.3 and 3.5\,mm. 
Although the improvement with stochastic optics is again evident from visual inspection, we can also quantify the improvement using standard fidelity metrics such as the (normalized) mean squared error (MSE):
\begin{align}
{\rm MSE} = \left. \left[\sum_{i,j} \left(I_{i,j} - I_{i,j}'\right)^2 \right]\middle/\left[\sum_{i,j} I_{i,j}^2 \right]\right.,
\end{align}
where $I_{i,j}$ is the model image and $I_{i,j}'$ is the reconstructed image. For the 3.5\,mm images in Figure~\ref{fig::Imaging_Comparisons}, the MSE is 0.188 for the direct reconstruction with MEM, is 0.189 for MEM with deblurring, and is 0.021 with stochastic optics. For the 1.3\,mm reconstruction, the MSE is 0.101 for direct MEM, is 0.092 for MEM with deblurring, and is 0.062 with stochastic optics.

Figure~\ref{fig::345_GHz_Example} shows a reconstruction at 345\,GHz (0.87\,mm). Although there is not yet active VLBI at 345\,GHz, this example illustrates the role of scattering mitigation on imaging when the scattering is minimal. We adopted the same source and array parameters as the 1.3\,mm example in Figure~\ref{fig::Imaging_Comparisons} (but note that some sites, such as the LMT and Pico Veleta do not currently have a 345\,GHz receiver, and realistic SEFDs will be significantly larger). In this case, the MSE is 0.034 for direct MEM, is 0.021 for MEM with deblurring, and is 0.018 with stochastic optics. In particular, deblurring and stochastic optics provide very similar results for this example because the array has no baselines that are significantly influenced by refractive substructure.

These examples have assumed that the reconstruction is done with complete knowledge of the time-averaged scattering statistics. Figure~\ref{fig::Varying_Alpha} shows reconstructions when the assumed value of the power-law exponent $\alpha$ is incorrect for stochastic optics reconstructions (the simulated image has $\alpha=5/3$). Because the refractive noise on a typical baseline $|\mathbf{b}| \sim r_0$ is approximately proportional to $(r_0/r_{\rm F})^{2-\alpha}$ \citep[see][]{Johnson_Gwinn_2015}, assuming lower/higher values of $\alpha$ will underestimate/overestimate the refractive noise. Thus, reconstructions that assume $\alpha=1.46$ have ``under-mitigated'' the scattering, while reconstructions that assume $\alpha=1.86$ have ``over-mitigated'' the scattering. Nevertheless, the image reconstructions are fairly robust to these significant errors in the assumed $\alpha$, and they still significantly outperform both deblurring and direct reconstructions without mitigation.

\section{Comparison with Existing Mitigation Strategies}
\label{sec::Discussion}

As the reconstructions in \S\ref{sec::Example_Reconstructions} demonstrate, stochastic optics provides effective scattering mitigation, even with only a single observing epoch. We now relate stochastic optics to existing scattering mitigation strategies. 

\subsection{Comparison with Adaptive Optics}
\label{sec::AO_Comparison}

We have already discussed some similarities and differences between stochastic optics and adaptive optics, but we now summarize and extend these comparisons:
\begin{itemize}
 \item {\bf Scattering Regime:} Adaptive optics mitigates observations in the weak-scattering regime, whereas stochastic optics mitigates observations in the strong-scattering regime. For imaging compact but resolved sources, the most significant difference between these regimes is that the dominant effect of strong scattering is image blurring, while the dominant effects of weak scattering are speckling and image wander. 
 \item {\bf Chromatic vs.\ Achromatic Scattering:} Optical scattering in the atmosphere is nearly achromatic \citep{Edlen_1966,TMS}, whereas radio scattering in the ISM is steeply chromatic (scattering angles are proportional to $\lambda^2$). Consequently, multi-frequency synthesis may significantly improve scattering mitigation for radio imaging.
 \item {\bf Tip-tilt:} VLBI observations are not usually sensitive to the absolute position of an image, which requires absolute phase referencing. Consequently, the lowest-order (and most important) correction of adaptive optics -- ``tip-tilt'' to correct for image wander -- is typically not required for VLBI. 
 \item {\bf Guide Stars:} Adaptive optics requires a bright point source (either a natural source or an artificial laser guide star) that is near the target (i.e., within the isoplanatic angle). Stochastic optics does not require a nearby point source and instead uses the known time-averaged power spectrum of the scattering to estimate the scattering screen via regularized minimization. Moreover, because the isoplanatic angle for refractive scintillation is equal to the scatter-broadening angle (see \S\ref{sec::Thin-Screen}), a scattered point source would have to be within the target scattered image to aid mitigation. For example, the Galactic Center magnetar, SGR~J1745-29, is $2.4''$ from \sgra\ \citep{Bower_2015} -- thousands of times the isoplanatic angle at millimeter wavelengths -- so cannot be used to mitigate the refractive scattering of \sgra. 
 \item {\bf Timescales:} At optical wavelengths, the coherence timescale is milliseconds, whereas the relevant coherence timescales for radio imaging range from a day to weeks. Consequently, radio-to-submillimeter observations only need to reconstruct a single scattering screen for an entire observing track, but single-epoch images are also sensitive to spurious image features from scattering. 
 \item {\bf Observables:} Adaptive optics dynamically perturbs the incident wavefront to correct images that are sampled directly using a filled aperture. Stochastic optics synthesizes images using a sparse set of measurements in the visibility domain.
 \item {\bf Implementation:} At optical wavelengths, wavefront corrections must be performed in real-time; radio interferometers coherently record the incoming signal so can perform scattering mitigation in post processing.
\end{itemize}
These comparisons show that stochastic optics has many advantages over adaptive optics (and other mitigation strategies at optical wavelengths), which help to overcome the lack of natural or artificial guide stars.

\subsection{Comparison with Deblurring and Deconvolution}
\label{sec::Deblurring} 

When the dominant measured effect of scattering is diffractive image blurring, simpler mitigation strategies such as ``deblurring'' and deconvolution can be effective \citep{Fish_2014}. These are motivated by the deterministic convolution of scattering in the ensemble-average scattering regime (see \S\ref{sec::Background}). Complex visibilities $V(\mathbf{b})$ of the scattered image are simply those of the unscattered image, multiplied by $\tilde{G}(\mathbf{b}) = e^{-\frac{1}{2}D_{\phi}(\mathbf{b}/(1+M))}$. In this regime, scattering mitigation is straightforward: one simply divides each measured visibility by $\tilde{G}(\mathbf{b})$ (and also divides the respective thermal noise by the same factor to preserve the signal-to-noise ratio) and then proceeds with standard synthesis imaging of the rescaled visibilities. After this correction, the action of scattering is only to amplify the thermal noise on long baselines. 

Although conceptually simple, deblurring has obvious limitations. As a specific example, VLBI of \sgra\ at 1.3\,cm has detected refractive noise of ${\sim}10\,{\rm mJy}$ on long baselines \citep{Gwinn_2014}. Deblurring these measurements would increase their correlated flux densities to higher amplitudes than the zero-baseline value, violating image positivity. The limitations of deblurring are less significant for the EHT because the expected refractive noise is only $1-2\%$ of the zero-baseline flux density \citep[${\sim}50\,{\rm mJy}$;][]{Johnson_Gwinn_2015} and the 1.3\,mm scattering kernel $\tilde{G}(\mathbf{b})$ never falls below ${\approx}0.2$ for EHT baselines \citep[see Figure~1 of][]{Fish_2014}. Consequently, deblurring will not produce unphysical images but will still tend to produce overly compact image reconstructions with artificial ``clumps'' of high flux to compensate for the substructure (see Figure~\ref{fig::Imaging_Comparisons}). Moreover, our imaging simulations suggest that deblurring should \underline{not} be used for imaging \sgra\ at wavelengths longer than 1.3\,mm. 

\citet{Fish_2014} has also discussed other image deconvolution approaches to mitigate scattering, such as the Wiener deconvolution, which could downweight long baseline measurements to account for their refractive noise. However, these methods must also be applied with caution because refractive scattering does not correspond to a convolution of the unscattered image with a position-independent kernel \citep[see][]{Johnson_Gwinn_2015}. Another problem with the Wiener deconvolution is that it assumes that noise is applied in the image domain, whereas thermal and systematic noise in VLBI is instead added in the visibility domain. Finally, the Wiener deconvolution assumes that the added noise is independent of the unscattered image, whereas refractive noise is highly sensitive to the unscattered image. 

Our method also differs from other approaches that approximate the scattering as a convolution, such as optical speckle interferometry \citep{Labeyrie_1970}. These approaches require that the scattering is isoplanatic with respect to the source, which for refractive scattering would require $\theta_{\rm src} \ll \theta_{\rm scatt}$ (see \S\ref{sec::Thin-Screen}). Because our primary cases of interest (and all the examples that we have shown) have $\theta_{\rm src} \gsim \theta_{\rm scatt}$, this assumption is not valid. 

Thus, while some of these alternative mitigation strategies may provide reasonable results, they are not formally consistent with the expected properties of refractive noise; consequently, a reduced data chi-squared near unity will not generally be expected, and the success of an imaging algorithm may be difficult to assess. In contrast, stochastic optics is consistent with the theoretical expectations for the scattering, it produces the most uninformative (as determined by the entropy function) unscattered image while accounting for the expected amount of image substructure from scattering, it produces unscattered images that are both ``deblurred'' and forced to be physically plausible (e.g., positive) by construction, and the resulting images will match the data to give a reduced chi-squared near unity. Even when deblurring produces acceptable results because refractive effects are minimal, stochastic optics should produce equivalent or superior reconstructions because the refractive power is regularized by the time-averaged power spectrum $Q(\mathbf{q})$ (see, e.g., Figure~\ref{fig::345_GHz_Example}).

\section{Summary}
\label{sec::Summary}

Interstellar scattering imposes a fundamental limitation on the resolution of VLBI. Without a scattering mitigation framework, increasing the observing sensitivity, bandwidth, or even the baseline coverage will not improve direct imaging beyond a limit set by scattering. And although scattering limitations are often irrelevant for VLBI, they are critical for imaging \sgra\ and other heavily scattered sources with ground-based VLBI, even at wavelengths as short as 1\,mm (see Figure~\ref{fig::Strehl_Ratio}). 

Stochastic optics can gracefully integrate the known limitations of scattering while also exploiting the many deterministic properties. For example, only one scattering screen is required to simultaneously image multiple frequencies, multiple nearby observing epochs, and all four Stokes parameters. Joint imaging at multiple frequencies could even allow a measurement of their relative image positions because the scattering screen provides a common point of reference. As a related example, by monitoring the total flux densities of intra-day variables at multiple frequencies, \citet{Macquart_2013} has estimated their frequency-dependent core shifts via relative lags caused by different reflex shifts of their diffraction patterns. 

The most important application of stochastic optics will be VLBI imaging of \sgra\ that includes ALMA, as is expected in 2017 \citep{Fish_2013}. Figure~\ref{fig::Imaging_Comparisons} shows that the simpler ``deblurring'' mitigation strategy developed by \citet{Fish_2014} may provide a good first approximation for EHT images of \sgra\ at 1.3~mm (and it should be an excellent strategy for the EHT at 0.87~mm; see Figure~\ref{fig::345_GHz_Example}), although the resulting images may tend to have too much compact structure. However, deblurring should \underline{not} be applied without modification at longer wavelengths, especially when sensitive sites such as ALMA, the GBT, and the LMT participate, since these will bring long-baseline detections that are heavily influenced (if not entirely dominated) by refractive noise \citep[see, e.g.,][]{Gwinn_2014}. Stochastic optics will be critical for synthesis imaging with these data and could also be applied for high-resolution imaging of other AGN with \emph{RadioAstron} at 18 and 6\,cm.

\acknowledgements{I am grateful to Andrew Chael for guidance in implementing stochastic optics in his VLBI imaging library. I am also indebted to Ramesh Narayan, Lindy Blackburn, Katie Bouman, and Vincent Fish for many invaluable conversations. I thank the National Science Foundation and the Gordon and Betty Moore Foundation (GBMF-3561) for financial support of this work. 
}

\ \\

\bibliography{Stochastic_Optics.bib}

\begin{thebibliography}{}
\expandafter\ifx\csname natexlab\endcsname\relax\def\natexlab#1{#1}\fi

\bibitem[{{Armstrong} {et~al.}(1995){Armstrong}, {Rickett}, \&
  {Spangler}}]{Armstrong_1995}
{Armstrong}, J.~W., {Rickett}, B.~J., \& {Spangler}, S.~R. 1995, \apj, 443, 209

\bibitem[{{Babcock}(1953)}]{Babcock_1953}
{Babcock}, H.~W. 1953, \pasp, 65, 229

\bibitem[{{Bardeen}(1973)}]{Bardeen_1973}
{Bardeen}, J.~M. 1973, Les Astres Occlus, 215

\bibitem[{{Blandford} \& {Narayan}(1985)}]{Blandford_Narayan_1985}
{Blandford}, R., \& {Narayan}, R. 1985, \mnras, 213, 591

\bibitem[{{Born} \& {Wolf}(1980)}]{Born_Wolf}
{Born}, M., \& {Wolf}, E. 1980, {Principles of Optics Electromagnetic Theory of
  Propagation, Interference and Diffraction of Light}

\bibitem[{Bouman {et~al.}(2016)Bouman, Johnson, Zoran, Fish, Doeleman, \&
  Freeman}]{Bouman_2016}
Bouman, K.~L., Johnson, M.~D., Zoran, D., {et~al.} 2016, in The IEEE Conference
  on Computer Vision and Pattern Recognition (CVPR)

\bibitem[{{Bower} {et~al.}(2006){Bower}, {Goss}, {Falcke}, {Backer}, \&
  {Lithwick}}]{Bower_2006}
{Bower}, G.~C., {Goss}, W.~M., {Falcke}, H., {Backer}, D.~C., \& {Lithwick}, Y.
  2006, \apjl, 648, L127

\bibitem[{{Bower} {et~al.}(2014){Bower}, {Deller}, {Demorest}, {Brunthaler},
  {Eatough}, {Falcke}, {Kramer}, {Lee}, \& {Spitler}}]{Bower_2014}
{Bower}, G.~C., {Deller}, A., {Demorest}, P., {et~al.} 2014, \apjl, 780, L2

\bibitem[{{Bower} {et~al.}(2015){Bower}, {Deller}, {Demorest}, {Brunthaler},
  {Falcke}, {Moscibrodzka}, {O'Leary}, {Eatough}, {Kramer}, {Lee}, {Spitler},
  {Desvignes}, {Rushton}, {Doeleman}, \& {Reid}}]{Bower_2015}
---. 2015, \apj, 798, 120

\bibitem[{{Brinkerink} {et~al.}(2016){Brinkerink}, {M{\"u}ller}, {Falcke},
  {Bower}, {Krichbaum}, {Castillo}, {Deller}, {Doeleman}, {Fraga-Encinas},
  {Goddi}, {Hern{\'a}ndez-G{\'o}mez}, {Hughes}, {Kramer}, {L{\'e}on-Tavares},
  {Loinard}, {Monta{\~n}a}, {Mo{\'s}cibrodzka}, {Ortiz-Le{\'o}n},
  {Sanchez-Arguelles}, {Tilanus}, {Wilson}, \& {Zensus}}]{Brinkerink_2016}
{Brinkerink}, C.~D., {M{\"u}ller}, C., {Falcke}, H., {et~al.} 2016, \mnras,
  462, 1382

\bibitem[{{Brisken} {et~al.}(2010){Brisken}, {Macquart}, {Gao}, {Rickett},
  {Coles}, {Deller}, {Tingay}, \& {West}}]{Brisken_2010}
{Brisken}, W.~F., {Macquart}, J.-P., {Gao}, J.~J., {et~al.} 2010, \apj, 708,
  232

\bibitem[{{Broderick} {et~al.}(2016){Broderick}, {Fish}, {Johnson},
  {Rosenfeld}, {Wang}, {Doeleman}, {Akiyama}, {Johannsen}, \&
  {Roy}}]{Broderick_2016}
{Broderick}, A.~E., {Fish}, V.~L., {Johnson}, M.~D., {et~al.} 2016, \apj, 820,
  137

\bibitem[{{Chael} {et~al.}(2016){Chael}, {Johnson}, {Narayan}, {Doeleman},
  {Wardle}, \& {Bouman}}]{Chael_2016}
{Chael}, A.~A., {Johnson}, M.~D., {Narayan}, R., {et~al.} 2016, \apj, 829, 11

\bibitem[{{Cordes} \& {Lazio}(2002)}]{NE2001}
{Cordes}, J.~M., \& {Lazio}, T.~J.~W. 2002, arXiv:astro-ph/0207156,
  astro-ph/0207156

\bibitem[{{Cornwell} \& {Evans}(1985)}]{Cornwell_Evans_1985}
{Cornwell}, T.~J., \& {Evans}, K.~F. 1985, \aap, 143, 77

\bibitem[{{Davies} \& {Kasper}(2012)}]{Davies_Kasper_2012}
{Davies}, R., \& {Kasper}, M. 2012, \araa, 50, 305

\bibitem[{{Doeleman} {et~al.}(2009){Doeleman}, {Agol}, {Backer}, {Baganoff},
  {Bower}, {Broderick}, {Fabian}, {Fish}, {Gammie}, {Ho}, {Honman},
  {Krichbaum}, {Loeb}, {Marrone}, {Reid}, {Rogers}, {Shapiro}, {Strittmatter},
  {Tilanus}, {Weintroub}, {Whitney}, {Wright}, \& {Ziurys}}]{Doeleman_2009}
{Doeleman}, S., {Agol}, E., {Backer}, D., {et~al.} 2009, in Astronomy, Vol.
  2010, astro2010: The Astronomy and Astrophysics Decadal Survey

\bibitem[{{Edl{\'e}n}(1966)}]{Edlen_1966}
{Edl{\'e}n}, B. 1966, Metrologia, 2, 71

\bibitem[{{Falcke} {et~al.}(2000){Falcke}, {Melia}, \& {Agol}}]{Falcke_2000}
{Falcke}, H., {Melia}, F., \& {Agol}, E. 2000, \apjl, 528, L13

\bibitem[{{Fish} {et~al.}(2013){Fish}, {Alef}, {Anderson}, {Asada}, {Baudry},
  {Broderick}, {Carilli}, {Colomer}, {Conway}, {Dexter}, {Doeleman}, {Eatough},
  {Falcke}, {Frey}, {Gab{\'a}nyi}, {G{\'a}lvan-Madrid}, {Gammie}, {Giroletti},
  {Goddi}, {G{\'o}mez}, {Hada}, {Hecht}, {Honma}, {Humphreys}, {Impellizzeri},
  {Johannsen}, {Jorstad}, {Kino}, {K{\"o}rding}, {Kramer}, {Krichbaum},
  {Kudryavtseva}, {Laing}, {Lazio}, {Loeb}, {Lu}, {Maccarone}, {Marscher},
  {Mart'{\i}-Vidal}, {Martins}, {Matthews}, {Menten}, {Miller}, {Miller-Jones},
  {Mirabel}, {Muller}, {Nagai}, {Nagar}, {Nakamura}, {Paragi}, {Pradel},
  {Psaltis}, {Ransom}, {Rodr'$\backslash$iguez}, {Rottmann}, {Rushton}, {Shen},
  {Smith}, {Stappers}, {Takahashi}, {Tarchi}, {Tilanus}, {Verbiest},
  {Vlemmings}, {Walker}, {Wardle}, {Wiik}, {Zackrisson}, \&
  {Zensus}}]{Fish_2013}
{Fish}, V., {Alef}, W., {Anderson}, J., {et~al.} 2013, ArXiv e-prints,
  arXiv:1309.3519

\bibitem[{{Fish} {et~al.}(2014){Fish}, {Johnson}, {Lu}, {Doeleman}, {Bouman},
  {Zoran}, {Freeman}, {Psaltis}, {Narayan}, {Pankratius}, {Broderick}, {Gwinn},
  \& {Vertatschitsch}}]{Fish_2014}
{Fish}, V.~L., {Johnson}, M.~D., {Lu}, R.-S., {et~al.} 2014, \apj, 795, 134

\bibitem[{{Fish} {et~al.}(2016){Fish}, {Johnson}, {Doeleman}, {Broderick},
  {Psaltis}, {Lu}, {Akiyama}, {Alef}, {Algaba}, {Asada}, {Beaudoin},
  {Bertarini}, {Blackburn}, {Blundell}, {Bower}, {Brinkerink}, {Cappallo},
  {Chael}, {Chamberlin}, {Chan}, {Crew}, {Dexter}, {Dexter}, {Dzib}, {Falcke},
  {Freund}, {Friberg}, {Greer}, {Gurwell}, {Ho}, {Honma}, {Inoue}, {Johannsen},
  {Kim}, {Krichbaum}, {Lamb}, {Le{\'o}n-Tavares}, {Loeb}, {Loinard},
  {MacMahon}, {Marrone}, {Moran}, {Mo{\'s}cibrodzka}, {Ortiz-Le{\'o}n},
  {Oyama}, {{\"O}zel}, {Plambeck}, {Pradel}, {Primiani}, {Rogers}, {Rosenfeld},
  {Rottmann}, {Roy}, {Ruszczyk}, {Smythe}, {SooHoo}, {Spilker}, {Stone},
  {Strittmatter}, {Tilanus}, {Titus}, {Vertatschitsch}, {Wagner}, {Wardle},
  {Weintroub}, {Woody}, {Wright}, {Yamaguchi}, {Young}, {Young}, {Zensus}, \&
  {Ziurys}}]{Fish_2016}
{Fish}, V.~L., {Johnson}, M.~D., {Doeleman}, S.~S., {et~al.} 2016, \apj, 820,
  90

\bibitem[{{Frieden}(1972)}]{Frieden_1972}
{Frieden}, B.~R. 1972, Journal of the Optical Society of America (1917-1983),
  62, 511

\bibitem[{{Goodman} \& {Narayan}(1989)}]{GoodmanNarayan89}
{Goodman}, J., \& {Narayan}, R. 1989, \mnras, 238, 995

\bibitem[{{Goodman} {et~al.}(1987){Goodman}, {Romani}, {Blandford}, \&
  {Narayan}}]{Goodman_1987}
{Goodman}, J.~J., {Romani}, R.~W., {Blandford}, R.~D., \& {Narayan}, R. 1987,
  \mnras, 229, 73

\bibitem[{{Gull} \& {Daniell}(1978)}]{Gull_Daniell_1978}
{Gull}, S.~F., \& {Daniell}, G.~J. 1978, \nat, 272, 686

\bibitem[{{Gwinn} {et~al.}(1998){Gwinn}, {Britton}, {Reynolds}, {Jauncey},
  {King}, {McCulloch}, {Lovell}, \& {Preston}}]{Gwinn_1998}
{Gwinn}, C.~R., {Britton}, M.~C., {Reynolds}, J.~E., {et~al.} 1998, \apj, 505,
  928

\bibitem[{{Gwinn} {et~al.}(2014){Gwinn}, {Kovalev}, {Johnson}, \&
  {Soglasnov}}]{Gwinn_2014}
{Gwinn}, C.~R., {Kovalev}, Y.~Y., {Johnson}, M.~D., \& {Soglasnov}, V.~A. 2014,
  \apjl, 794, L14

\bibitem[{{H{\"o}gbom}(1974)}]{Hogbom_1974}
{H{\"o}gbom}, J.~A. 1974, \aaps, 15, 417

\bibitem[{Jackson(1999)}]{Jackson_1999}
Jackson, J.~D. 1999, Classical electrodynamics, 3rd edn. (New York, {NY}:
  Wiley)

\bibitem[{{Johnson} \& {Gwinn}(2015)}]{Johnson_Gwinn_2015}
{Johnson}, M.~D., \& {Gwinn}, C.~R. 2015, \apj, 805, 180

\bibitem[{{Johnson} \& {Narayan}(2016)}]{Johnson_Narayan_2016}
{Johnson}, M.~D., \& {Narayan}, R. 2016, \apj, 826, 170

\bibitem[{{Johnson} {et~al.}(2016){Johnson}, {Kovalev}, {Gwinn}, {Gurvits},
  {Narayan}, {Macquart}, {Jauncey}, {Voitsik}, {Anderson}, {Sokolovsky}, \&
  {Lisakov}}]{Johnson_2016}
{Johnson}, M.~D., {Kovalev}, Y.~Y., {Gwinn}, C.~R., {et~al.} 2016, \apjl, 820,
  L10

\bibitem[{{Kardashev} {et~al.}(2013){Kardashev}, {Khartov}, {Abramov},
  {Avdeev}, {Alakoz}, {Aleksandrov}, {Ananthakrishnan}, {Andreyanov},
  {Andrianov}, {Antonov}, {Artyukhov}, {Arkhipov}, {Baan}, {Babakin},
  {Babyshkin}, {Bartel'}, {Belousov}, {Belyaev}, {Berulis}, {Burke},
  {Biryukov}, {Bubnov}, {Burgin}, {Busca}, {Bykadorov}, {Bychkova},
  {Vasil'kov}, {Wellington}, {Vinogradov}, {Wietfeldt}, {Voitsik},
  {Gvamichava}, {Girin}, {Gurvits}, {Dagkesamanskii}, {D'Addario},
  {Giovannini}, {Jauncey}, {Dewdney}, {D'yakov}, {Zharov}, {Zhuravlev},
  {Zaslavskii}, {Zakhvatkin}, {Zinov'ev}, {Ilinen}, {Ipatov}, {Kanevskii},
  {Knorin}, {Casse}, {Kellermann}, {Kovalev}, {Kovalev}, {Kovalenko}, {Kogan},
  {Komaev}, {Konovalenko}, {Kopelyanskii}, {Korneev}, {Kostenko}, {Kotik},
  {Kreisman}, {Kukushkin}, {Kulishenko}, {Cooper}, {Kut'kin}, {Cannon},
  {Larionov}, {Lisakov}, {Litvinenko}, {Likhachev}, {Likhacheva}, {Lobanov},
  {Logvinenko}, {Langston}, {McCracken}, {Medvedev}, {Melekhin}, {Menderov},
  {Murphy}, {Mizyakina}, {Mozgovoi}, {Nikolaev}, {Novikov}, {Novikov},
  {Oreshko}, {Pavlenko}, {Pashchenko}, {Ponomarev}, {Popov}, {Pravin-Kumar},
  {Preston}, {Pyshnov}, {Rakhimov}, {Rozhkov}, {Romney}, {Rocha}, {Rudakov},
  {R{\"a}is{\"a}nen}, {Sazankov}, {Sakharov}, {Semenov}, {Serebrennikov},
  {Schilizzi}, {Skulachev}, {Slysh}, {Smirnov}, {Smith}, {Soglasnov},
  {Sokolovskii}, {Sondaar}, {Stepan'yants}, {Turygin}, {Turygin}, {Tuchin},
  {Urpo}, {Fedorchuk}, {Finkel'shtein}, {Fomalont}, {Fejes}, {Fomina},
  {Khapin}, {Tsarevskii}, {Zensus}, {Chuprikov}, {Shatskaya}, {Shapirovskaya},
  {Sheikhet}, {Shirshakov}, {Schmidt}, {Shnyreva}, {Shpilevskii}, {Ekers}, \&
  {Yakimov}}]{Kardashev_2013}
{Kardashev}, N.~S., {Khartov}, V.~V., {Abramov}, V.~V., {et~al.} 2013,
  Astronomy Reports, 57, 153

\bibitem[{{Labeyrie}(1970)}]{Labeyrie_1970}
{Labeyrie}, A. 1970, \aap, 6, 85

\bibitem[{{Lu} {et~al.}(2011){Lu}, {Krichbaum}, {Eckart}, {K{\"o}nig},
  {Kunneriath}, {Witzel}, {Witzel}, \& {Zensus}}]{Lu_2011}
{Lu}, R.-S., {Krichbaum}, T.~P., {Eckart}, A., {et~al.} 2011, \aap, 525, A76

\bibitem[{{Luminet}(1979)}]{Luminet_1979}
{Luminet}, J.-P. 1979, \aap, 75, 228

\bibitem[{{Macquart} {et~al.}(2013){Macquart}, {Godfrey}, {Bignall}, \&
  {Hodgson}}]{Macquart_2013}
{Macquart}, J.-P., {Godfrey}, L.~E.~H., {Bignall}, H.~E., \& {Hodgson}, J.~A.
  2013, \apj, 765, 142

\bibitem[{{Narayan}(1992)}]{Narayan_1992}
{Narayan}, R. 1992, Royal Society of London Philosophical Transactions Series
  A, 341, 151

\bibitem[{{Narayan} \& {Goodman}(1989)}]{NarayanGoodman89}
{Narayan}, R., \& {Goodman}, J. 1989, \mnras, 238, 963

\bibitem[{{Narayan} \& {Nityananda}(1986)}]{Narayan_Nityananda_1986}
{Narayan}, R., \& {Nityananda}, R. 1986, \araa, 24, 127

\bibitem[{{Ortiz-Le{\'o}n} {et~al.}(2016){Ortiz-Le{\'o}n}, {Johnson},
  {Doeleman}, {Blackburn}, {Fish}, {Loinard}, {Reid}, {Castillo}, {Chael},
  {Hern{\'a}ndez-G{\'o}mez}, {Hughes}, {Le{\'o}n-Tavares}, {Lu}, {Monta{\~n}a},
  {Narayanan}, {Rosenfeld}, {S{\'a}nchez}, {Schloerb}, {Shen}, {Shiokawa},
  {SooHoo}, \& {Vertatschitsch}}]{Ortiz_2016}
{Ortiz-Le{\'o}n}, G.~N., {Johnson}, M.~D., {Doeleman}, S.~S., {et~al.} 2016,
  \apj, 824, 40

\bibitem[{{Pearson} \& {Readhead}(1984)}]{Pearson_Readhead_1984}
{Pearson}, T.~J., \& {Readhead}, A.~C.~S. 1984, \araa, 22, 97

\bibitem[{{Rickett}(1990)}]{Rickett_1990}
{Rickett}, B.~J. 1990, \araa, 28, 561

\bibitem[{{Rickett} {et~al.}(1984){Rickett}, {Coles}, \& {Bourgois}}]{RCB_1984}
{Rickett}, B.~J., {Coles}, W.~A., \& {Bourgois}, G. 1984, \aap, 134, 390

\bibitem[{{Taylor}(1938)}]{Taylor_1938}
{Taylor}, G.~I. 1938, Proceedings of the Royal Society of London Series A, 164,
  476

\bibitem[{{Thi{\'e}baut}(2013)}]{Thiebaut_2013}
{Thi{\'e}baut}, {\'E}. 2013, in EAS Publications Series, Vol.~59, EAS
  Publications Series, ed. D.~{Mary}, C.~{Theys}, \& C.~{Aime}, 157--187

\bibitem[{{Thompson} {et~al.}(2001){Thompson}, {Moran}, \& {Swenson}}]{TMS}
{Thompson}, A.~R., {Moran}, J.~M., \& {Swenson}, Jr., G.~W. 2001,
  {Interferometry and Synthesis in Radio Astronomy, 2nd Edition}

\bibitem[{{Walker} {et~al.}(2008){Walker}, {Koopmans}, {Stinebring}, \& {van
  Straten}}]{Walker_2008}
{Walker}, M.~A., {Koopmans}, L.~V.~E., {Stinebring}, D.~R., \& {van Straten},
  W. 2008, \mnras, 388, 1214

\end{thebibliography}

\end{document}